\documentclass[%
headsepline=false,
abstract=true,
numbers=noendperiod
]{scrartcl}
\usepackage{comment} 
\usepackage[UKenglish]{babel}                       
\usepackage[utf8]{inputenc}							
\usepackage[T1]{fontenc}							
\usepackage{lmodern}                                
\usepackage{fourier}								
\usepackage[section]{placeins}					    
\usepackage{booktabs}								
\usepackage{csquotes}                               
\usepackage{mathtools}								
\usepackage{amsmath}                                
\usepackage{amssymb}								
\usepackage{tensor}                                 
\usepackage{dsfont}                                 
\usepackage[dvipsnames]{xcolor}                     
\usepackage[mathscr]{euscript}					    
\usepackage{xfrac}                                  
\usepackage{tensor}                                 
\usepackage{stmaryrd}                               
\usepackage[%
backend=biber,
style=numeric,
citestyle=numeric-comp,
sorting=none,
doi=false,
isbn=false,
url=true,
maxbibnames=99
]{biblatex}                                         
\usepackage[pdfborder={0 0 0}]{hyperref}			
\usepackage{microtype}                              
\usepackage{tikz-cd} 
\usepackage[authormarkup=none,defaultcolor=RedOrange,commandnameprefix=ifneeded]{changes}
\newcommand{\e}[1]{\operatorname{e}^{#1}}
\renewcommand{\i}{\operatorname{i}\!}
\renewcommand{\d}{\operatorname{d}\!}
\newcommand{\tr}{\operatorname{tr}}

\newcommand{\sl}{\mathfrak{sl}}
\newcommand{\g}{\mathfrak{g}}
\newcommand{\iso}{\mathfrak{iso}}
\newcommand{\so}{\mathfrak{so}}
\newcommand{\hs}{\mathfrak{hs}}
\newcommand{\ihs}{\mathfrak{ihs}}
\newcommand{\dsR}{\mathds{R}}
\newcommand{\dsN}{\mathds{N}}
\newcommand{\dsZ}{\mathds{Z}}
\newcommand{\scrim}{M}

\newcommand{\idty}{\mathds{1}}
\newcommand{\ad}{\operatorname{ad}}
\newcommand{\tad}[1]{\tensor*[^{\tau}]{\operatorname{ad}}{#1}}
\newcommand{\ead}[1]{\tensor*[^{\star}]{\operatorname{ad}}{#1}}
\newcommand{\bldidty}{\boldsymbol{1}}
\newcommand{\bldtau}{\boldsymbol{\tau}}
\DeclareMathAlphabet{\mathpzc}{OT1}{pzc}{m}{it}     
\numberwithin{equation}{section}
\KOMAoptions{parskip=never}
\renewbibmacro{in:}{}   
\begin{document}
\title{Higher-Spin Gravity in Two Dimensions\\[0.2cm] with Vanishing Cosmological Constant}
\author{Xavier Bekaert\textsuperscript{a}\thanks{\href{mailto:xavier.bekaert@univ-tours.fr}{xavier.bekaert@univ-tours.fr}} \and Michel Pannier\textsuperscript{b,c}\thanks{\href{mailto:michel.pannier@unina.it}{michel.pannier@unina.it}}}
\publishers{%
\vspace{0.6cm}
\textsuperscript{\footnotesize a\ }\begin{minipage}[t]{0.8\linewidth}
    {\footnotesize \emph{Institut Denis Poisson, Unité Mixte de Recherche 7013 du CNRS,\\
    Université de Tours \& Université d'Orléans,\\
    Parc de Grandmount, F-37200 Tours, France} \par}
    \end{minipage}\\[0.1cm]
\textsuperscript{\footnotesize b\ }\begin{minipage}[t]{0.8\linewidth}
    {\footnotesize \emph{Dipartimento di Fisica ``Ettore Pancini'', Università degli Studi di Napoli ``Federico II''\\ Via Cintia, 21, I-80126 Napoli, Italy} \par}
    \end{minipage}\\[0.1cm]
\textsuperscript{\footnotesize c\ }\begin{minipage}[t]{0.8\linewidth}
    {\footnotesize \emph{Istituto Nazionale di Fisica Nucleare (INFN), Sezione di Napoli\\ Via Cintia, 21, I-80126 Napoli, Italy} \par}
    \end{minipage}\\[0.1cm]
}%
\date{}
\maketitle
\thispagestyle{empty}
\setcounter{page}{0}
\begin{abstract}
In this paper, we use a version of the BF formulation of two-dimensional dilaton gravity that allows to define a gauge theory of the two-dimensional Poincaré or Maxwell algebras and several of their higher-spin generalisations, both of finite and infinite dimension. The spectrum of the two-dimensional higher-spin gravity with vanishing cosmological constant based on the extended, infinite-dimensional higher-spin algebra is shown to contain an infinite collection of scalar degrees of freedom with a continuum of ever increasing mass, corresponding to the twisted-(co)adjoint representation. We comment on an approach to include backreaction of the scalar fields on the gravity sector at the level of formal equations of motion, thereby providing a first example of a fully interacting higher-spin gravity theory with vanishing cosmological constant in two dimensions.
\end{abstract}
\pagebreak
\tableofcontents
\vspace{0.5cm}
\section{Introduction}\label{sec:intro}
The exploration of higher-spin symmetry is interesting for a variety of reasons, notably because integrable models typically exhibit deformed higher-spin symmetries (while free models exhibit linear higher-spin symmetries). Applying the holographic dictionary to the Noe\-ther currents associated to higher-spin symmetries implies that the holographic duals to free or integrable conformal field theories must be higher-spin gravity theories. Such  holographic dualities are typically easier to analyse since both sides (boundary correlators and bulk interactions) are strongly restricted by a huge symmetry group. Ideally, the hope is to find cases in which both sides of the duality are fully solvable, eventually allowing us to prove the duality or, at least, learn more about holography itself. For instance, higher-spin gravity in a bulk spacetime with negative cosmological constant should be dual to $\mathit{O}(N)$ vector models for bulk spacetime dimensions $d\geqslant 4$ \cite{Klebanov:2002ja,Sezgin:2003pt,Leigh:2003gk,Giombi:2011kc}, and to $\mathcal{W}_N$- or $\mathcal{W}_\infty$-minimal models for a three-dimensional bulk \cite{Campoleoni:2010zq,Henneaux:2010xg,Gaberdiel:2012uj}, within a semi-classical limit. As argued in \cite{Alkalaev:2019xuv,Anninos:2023lin,Bekaert:2025azj}, applying similar lines of reasoning to the (A)dS$_2$/CFT$_1$ correspondence suggests that some Sachdev–Ye–Kitaev (SYK) models exhibiting higher-spin symmetry should be dual to a higher-spin extension of Jackiw-Teitelboim (JT) gravity \cite{Teitelboim:1983ux,Jackiw:1984je}. More precisely, this extension must include a suitable matter sector, i.e. an infinite tower of scalar bulk fields dual to singlet bilinear boundary operators \cite {Gross:2017vhb,Gross:2017aos}.

Fortunately, the task of constructing consistent, possibly interacting, theories of higher-spin gravity becomes much friendlier in the lower-dimensional cases (i.e. $d=2$ and $d=3$) when all gauge fields are topological. In this case, one actually has the luxury of an action principle, namely Chern-Simons gauge theories in three dimensions and BF (or, more generally, Poisson-Sigma) models in two dimensions. At non-vanishing cosmological constant $\Lambda\neq 0$, both cases have been generalised into higher-spin gravity theories, containing either a finite or infinite number of gauge fields \cite{Blencowe:1988gj,Alkalaev:2013fsa,Grumiller:2013swa,Alkalaev:2014qpa}. A dynamical sector in the form of massive matter fields can then be coupled to the topological theory on different levels: in three dimensions, a pair of additional complex scalar fields of fine-tuned mass is described on-shell in terms of an unfolded equation \cite{Prokushkin:1998bq,Prokushkin:1998vn} (and these turn out to be holographically dual to primary operators of the dual minimal-model CFT \cite{Gaberdiel:2010pz}); in two dimensions, the matter sector can be directly introduced at the level of the (BF-type) action and contains an infinite tower of scalars with increasing mass \cite{Alkalaev:2019xuv,Alkalaev:2020kut}. The spectrum of the latter, two-dimensional theory includes local degrees of freedom although the field equations take the form of flatness and covariant-constancy conditions, because the fields take values in a suitable extension of the \textit{infinite-dimensional} higher-spin algebra. There exists another proposal of two-dimensional higher-spin gravity \cite{Vasiliev:1995sv}, the matter spectrum of which is made of a single scalar field with a fixed mass.

For the case of vanishing cosmological constant, such theories are not quite as well explored, which is to be contrasted with the importance of locally flat spacetimes from the viewpoint of bulk physics: these are clearly a good approximation to real-world settings, whereas $\Lambda<0$ does not seem to be realised in our universe. Furthermore, theories around Minkowski spacetime $\dsR^{1,1}$ easily evade the no-go theorems \cite{Antunes:2025iaw} against higher-spin charges in AdS$_2$. Previous developments in the three-dimensional flat case include the introduction of finite-spin generalisations \cite{Afshar:2013vka,Gonzalez:2013oaa}, proposals of infinite-dimensional higher-spin algebras from \.{I}nönü-Wigner contractions \cite{Grumiller:2014lna,Ammon:2017vwt}, as well as universal-enveloping-algebra constructions \cite{Ammon:2020fxs,Campoleoni:2021blr}; furthermore, an unfolded description of massive fields has been proposed, both for scalars \cite{Ammon:2020fxs} and arbitrary-spin fields \cite{Ammon:2022vjr}. Note that, in two dimensions massive fields are topological, with the exception of scalar (and spinor) fields. 

In the present work, we introduce a higher-spin generalisation of JT gravity in the case of $\Lambda=0$, which will take the form of a BF-type theory that contains both the topological higher-spin gauge fields and the propagating matter sector. The usual incarnation of BF theory is in terms of a one-form $A$ and a zero-form $B$, both taking values in the adjoint representation of a Lie algebra $\g$ which is \textit{quadratic}, i.e. admitting a non-degenerate invariant symmetric bilinear form. The latter condition is guaranteed when $\g$ is a finite-dimensional semisimple Lie algebra since those algebras are endowed with the Killing form. However, this is not guaranteed when the Lie algebra admits an Abelian ideal, as it is the case for \.{I}nönü-Wigner contractions. For instance, in the flat limit the Killing form of the AdS$_2$ isometry algebra $\so(2,1)$ becomes degenerate and, in fact, there does not exist any bilinear form on the Poincaré algebra $\iso(1,1)$ which is both non-degenerate and adjoint-invariant.

Fortunately, there is a simple way out. A more general version of BF theory is formulated in terms of a one-form $A$ and a zero-form $B^*$ taking values in dual representations: the adjoint and the coadjoint representation, respectively. These two formulations of BF theory are equivalent for quadratic algebras $\g$, but the latter is more general. This solves the previous problem because the coadjoint representation actually admits a smooth limit for any \.{I}nönü-Wigner contraction (this is particularly manifest if one expresses this representation in terms of the structure constants). In this way, one can perform the \.{I}nönü-Wigner contraction of any BF theory. In particular, this was performed for the gauge formulation of JT gravity in order to obtain its flat limit \cite{Jackiw:1992bw}, as well as its various (non-relativistic and ultra-relativistic) limits \cite{Grumiller:2020elf}. This will be our strategy for the extension of JT gravity extended by a tower of higher-spin gauge fields and scalar matter fields.

In Section~\ref{sec:review}, we revisit the case $\Lambda\ne 0$ from the viewpoint of the pairing between the adjoint and coadjoint representations. More precisely, it will be introduced for the standard example of the gauge formulation of JT gravity in Subsection \ref{subsec:classical_grav}; then both finite-dimensional and infinite-dimensional higher-spin generalisations, as well as the extended version of the latter, are briefly discussed in Subsection \ref{subsec:higher-spin_BF}. Afterwards, in Subsection~\ref{subsec:tw_basis_spectrum} we present a particular approach to decompose the higher-spin algebra $\hs[\lambda]$, on which $\so(2,1)$ acts both through the adjoint and the twisted-adjoint representation, into irreducible sub-modules (known as \emph{Killing} and \emph{Weyl modules}, respectively) by explicitly defining adapted bases in both cases. In the twisted-adjoint basis, this results in the discrete mass spectrum of the matter sector.

From the particular manner in which we revisit the curved case, it will turn out rather straightforward to generalise to the flat case in Section~\ref{sec:flat_case}. We define the flat-space higher-spin algebra $\ihs[\scrim]$ as well as an adjoint and a twisted-adjoint basis (of a suitable vector-space completion, $\widehat{\ihs}[\scrim]$) in Subsection~\ref{subsec:ihs_tw_basis}, which allows us to decompose the twisted-adjoint representation of $\iso(1,1)$ on $\widehat{\ihs}[\scrim]$ into irreducible submodules and, consequently, present an explicit parametrisation of the masses of the scalar fields in Subsection~\ref{subsec:flat_BF_theory}. The spectrum of masses turns out to be continuous. Finally, we present the fundamental ingredients to a fully interacting higher-spin theory in Subsection~\ref{subsec:backreaction}.

The Cangemi-Jackiw (CJ) gravity theory \cite{Cangemi:1992bj} corresponds to the BF action for the Maxwell algebra in two dimensions and is known to be a reformulation of a conformally transformed version of the matterless Callan–Giddings–Harvey–Strominger (CGHS) model \cite{Callan:1992rs}, sometimes denoted $\widehat{\text{CGHS}}$ model \cite{Afshar:2019axx}. In Section~\ref{sec:higher-spin_CGHS}, a higher-spin extension is proposed by noticing that the Weyl algebra provides a natural infinite-dimensional extension of the Maxwell algebra. By taking inspiration from standard algebraic techniques in higher-spin gravity (see e.g. the review \cite{Vasiliev:1995dn}), one considers a doubled Weyl algebra and takes advantage of the existence of a trace and of an associative deformation, respectively, to propose a standard BF action for a higher-spin extension of Cangemi-Jackiw gravity and to deduce the existence of a deformed theory with cross-couplings between gauge and matter sectors.

The concluding section, which contains a brief summary and an outlook on a number of possible future directions, can be found in Section~\ref{sec:discussion}. Within two appendices we give some comments on the relation between bilinear forms and dual pairings (in Appendix~\ref{app:bilinear_pairing}) as well as details on the construction of solutions to a certain algebraic equation (in Appendix~\ref{app:sec:systematic_app}).
\section{Revisiting Jackiw-Teitelboim Gravity and its Generalisations}\label{sec:review}
In this section, we give a brief discussion of two-dimensional gravity and its formulation as a BF gauge theory, both for the case of negative and vanishing cosmological constant $\Lambda$. In the latter case, we replace the trace in the BF action by a pairing between the Lie algebra and its dual; this formulation will subsequently be used to revisit its higher-spin generalisation in the case of $\Lambda<0$.
\subsection{Gauge Theory of Two-Dimensional Gravity}\label{subsec:classical_grav}
The case of spacetime dimension two is special since the Einstein tensor vanishes identically and the Einstein-Hilbert action is a topological invariant. Nevertheless, following the proposal of Jackiw and Teitelboim \cite{Teitelboim:1983ux,Jackiw:1984je} (see e.g. \cite{Jackiw:1992bw} for an enlightening review) a theory of gravitation may be formulated, where the equation of motion for the Ricci scalar, $R-2\Lambda=0$, is realised as a consequence of an action principle at the price of introducing a dilaton field.

Similar to the case of three dimensions, in which gravity can be formulated as a Chern-Simons gauge theory of the respective spacetime symmetry algebra, the two-dimensional theory allows a gauge-theoretic formulation, as well \cite{Fukuyama:1985gg,Isler:1989hq}. In particular, one may consider a one-form $A=\varvarpi J+e^a P_a$  taking values in the isometry algebra of the respective vacuum spacetime, $\so(2,1)$ in the case of negative (or positive\footnote{In the algebraic formulation that we are following, we only make use of the isometry algebras of AdS$_2$ and dS$_2$ which are isomorphic, so they are not distinguished at this preliminary level where global issues are not considered. For brevity, we will always refer to the case of negative cosmological constant, only.}) cosmological constant, $\iso(1,1)$ at vanishing cosmological constant:\footnote{The metric convention here is $\eta=\operatorname{diag}(-1,1)$ and the Levi-Civita tensor is normalised as $\varepsilon_{01}=1$. In the $\sl(2,\dsR)$-like basis where $P_\pm=P_1\pm P_0$, it is $[J\,,P_\pm]=\pm P_\pm$ and $[P_+\,,P_-]=-2\Lambda J$.}
\begin{align}\label{eq:algebra_generalL}
    \left[J\,,P_a\right]=\varepsilon_a{}^b P_b\,, && \left[P_a\,,P_b\right]=-\Lambda\varepsilon_{ab}J\,.
\end{align}
The equation of motion is then given as the vanishing of the gauge curvature $F=\d A+A\wedge A$ and in the case of $\Lambda\ne 0$ can be derived by variation of the so-called BF action \cite{Isler:1989hq,Chamseddine:1989wn}
\begin{align}\label{eq:BF_action_tr}
    S[A,B]=\int\tr\left(B\cdot F\right)\,,
\end{align}
where $A$ is a one-form and $B$ is a zero-form, both taking values in the defining representation of $\sl(2,\dsR)\simeq\so(2,1)$ and ``\,$\cdot$\,'' denotes the associative matrix product. Since we are only interested in the equations of motion, we discard the possibility of any boundary term. The equation of motion for the algebra-valued zero-form $B$ (that describes the dilaton field) reads $\operatorname{D}\!B\equiv\d B+\ad_A(B)=0$, where the adjoint action is simply the commutator, $\ad_A(B)=[A\,,B]$. A first-order form of Jackiw-Teitelboim (JT) gravity à la Cartan \cite{Fukuyama:1985gg} is obtained from \eqref{eq:BF_action_tr} by decomposing all fields into Lorentz and translation generators (e.g. $A=\varvarpi J+e^a P_a$).

If no particular representation is chosen for the isometry algebra, the trace operation ``tr'' in \eqref{eq:BF_action_tr} refers instead to an adjoint-invariant bilinear form $\langle\,.\,,\,.\,\rangle$ on the algebra, meaning that $\tr(X\cdot Y)=\langle X\,,Y\rangle$, which in the case of $\g=\so(2,1)$ has non-vanishing components $\langle J\,,J\rangle=1$ and $\langle P_a\,,P_b\rangle=-\Lambda\eta_{ab}$. The slightly more general\footnote{Note that for a BF action principle of the form \eqref{eq:BF_action_bil}, it is not necessary for the bilinear form to be symmetric.} form of the BF action now reads
\begin{align}\label{eq:BF_action_bil}
    S[A,B]=\int\langle B\,,F\rangle\,.
\end{align}
Herein lies a first peculiarity of the flat-space case: there exists no adjoint-invariant, non-degenerate, bilinear form on the Poincaré algebra $\iso(1,1)$.\footnote{To actually see that any adjoint-invariant bilinear form on $\iso(1,1)$ is necessarily degenerate on the translation subalgebra $\dsR^2$: firstly, take $X=P_a$, $Y=P_b$ and $Z=J$ in the condition \eqref{eq:ad_inv} to deduce $\langle P_a,P_b\rangle=0$; secondly, set instead $X=P_a$ and $Y=Z=J$ in \eqref{eq:ad_inv} to obtain $\langle P_a,J\rangle=0$.} Remember that the adjoint-invariance means that the identity
\begin{align}\label{eq:ad_inv}
    \big\langle\,[X\,,Y]\,,Z\,\big\rangle=\big\langle\, X\,,[Y\,,Z]\,\big\rangle
\end{align}
is fulfilled for any triple $X,Y,Z\in\g$. The non-degeneracy and adjoint-invariance are in tension with each other when $\g$ possesses a non-trivial Abelian ideal.

At this point we would like to recall a refined, more flexible formulation of the BF action. Let $\g$ be the Lie algebra that the one-form $A$ takes values in, and let $\g^*$ be the respective dual space. Naturally, there is a pairing $(\,.\,,\,.\,):\,\g^*\times\g\rightarrow\dsR$ that produces the value of a dual element on an algebra element and this pairing is non-degenerate. If the Lie algebra $\g$ is endowed with a non-degenerate, adjoint-invariant, bilinear form (e.g. the Killing form when $\g$ is a finite-dimensional semisimple Lie algebra) then there exists an injective linear mapping 
\begin{align}
    \Phi:\ \g\rightarrow\g^*:\ X\mapsto\Phi\cdot X\,,
\end{align}
defined by $\left(\Phi\cdot X\,,Y\right)=\left\langle X\,,Y\right\rangle$. On the one hand, the invariance of the bilinear form translates into the equivariance of the linear map, more precisely $\Phi$ intertwines the adjoint and coadjoint representation, i.e. $\Phi\circ\ad=\ad^*\!\circ\,\Phi$, where the coadjoint action is defined in terms of the pairing in the usual way, i.e.
\begin{align}\label{eq:def_coadj_action}
    \left(\ad^*_X(Y^*)\,,Y\right):=-\left(Y^*,\ad_X(Y)\right)\,, && X,Y\in\g\,,\ \ Y^*\in\g^*\,.
\end{align}
Conversely, the existence of an injective linear mapping $\Phi:\,\g\rightarrow\g^*$ obeying this property allows to define a non-degenerate bilinear form on $\g$ via $\left\langle X\,,Y\right\rangle:=\left(\Phi\cdot X\,,Y\right)$. This bilinear form is adjoint-invariant if and only if the linear map is equivariant.\footnote{Note that even if the bilinear form is not invariant, nevertheless it fulfils a similar identity (see e.g. \cite[app.\,2.B]{Arnold89}).} In Appendix~\ref{app:bilinear_pairing}, we consider this definition for the cases $\g=\so(2,1)$ and $\g=\iso(1,1)$. 

For the present purpose, the important insight is that, in the case of the algebra $\g=\so(2,1)$, one can re-write the BF action \eqref{eq:BF_action_tr} in terms of the pairing between the algebra and its dual,
\begin{align}\label{eq:BF_action_pairing}
    \boxed{%
    S[A,B^*]=\int\left(B^*,F\right)\,,
    }%
\end{align}
without changing the physical content of the theory. But now, this new formulation allows the case $\iso(1,1)$ as well. In fact, the action \eqref{eq:BF_action_pairing} defines a BF theory which includes the case of vanishing cosmological constant \cite{Jackiw:1992bw}. In this more general formulation, the dilaton is described by a zero-form $B^*\in\g^*$, which can be identified with the image of $B$ under $\Phi$ when $\Lambda\neq 0$ (in which case one could still denote the action as a functional of $B$). 

On the one hand, the equation of motion resulting from the variation of $B^*$ in \eqref{eq:BF_action_pairing} remains, as before, the flatness of the gauge curvature. On the other hand, the equation of motion resulting from the variation of $A$ reads in terms of the coadjoint action as
\begin{align}\label{eq:eom_dualB}
    \d B^*+\ad^*_A(B^*)=0\,.
\end{align}
Note that the BF action is invariant under gauge transformations when $F$ transforms under the adjoint action and $B^*$ under the coadjoint action.\footnote{To be precise, from the transformation behaviour of the gauge-field one-form, $A\mapsto gAg^{-1}+g\d g^{-1}$, follows that the field strength transforms under the adjoint action of the group, i.e. $F\mapsto \operatorname{Ad}_g(F)=gFg^{-1}$. The dilaton zero-form simultaneously transforms like $B^*\mapsto\operatorname{Ad}^*_g(B^*)$, where the coadjoint group action is defined via the pairing, $(\operatorname{Ad}^*_g(X^*)\,,X):=(X^*,\operatorname{Ad}_{g^{-1}}(X))$, which immediately implies the invariance property.}
\subsection{Higher-Spin Generalisations}\label{subsec:higher-spin_BF}
The higher-spin algebras relevant in two dimensions when $\Lambda\neq 0$ can be subdivided into three categories of higher-spin extensions of $\so(2,1)\simeq\sl(2,\dsR)$: the finite-dimensional algebras $\sl(N,\dsR)$ leading to a finite tower of topological gauge fields, the infinite-dimensional algebras $\hs[\lambda]$ leading to an infinite tower of gauge fields, and finally the extended algebras $\hs[\lambda]\rtimes\dsZ_2$ leading to an infinite tower of gauge and matter fields.

As will be mentioned below for the finite-dimensional case and in Section~\ref{sec:flat_case} for the infinite-dimensional case, these higher-spin algebras admit analogues at vanishing cosmological constant, i.e. higher-spin extensions of $\iso(1,1)$, which can be defined intrinsically for $\Lambda=0$ or seen as \.{I}nönü-Wigner contractions $\Lambda\to0$.
\minisec{Finite-Dimensional Higher-Spin Algebras}
The simplest generalisation of BF theory to higher-spin gravity was performed for the case $\Lambda<0$ in \cite{Alkalaev:2013fsa,Grumiller:2013swa} and amounts to introducing a finite number of topological gauge fields and dilaton-like fields by replacing the $\sl(2,\dsR)$ algebra by $\sl(N,\dsR)$. Under the principal embedding $\sl(2,\dsR)\subset\sl(N,\dsR)$ the adjoint representation decomposes as
\begin{align}
    \sl(N,\dsR) = \bigoplus_{j=1}^{N-1} \mathcal{D}_j\,,
\end{align}
where $\mathcal{D}_j$ denotes the spin-$j$ irreducible representation of $\sl(2,\dsR)$. This decomposition gives rise to generators $V^{j+1}_m$ with $m = -j, \dots, j$, corresponding to topological gauge fields of spin $s=j+1$. In particular, $V^2_0=J$ and $V^2_{\pm1}=P_\pm$.

The \.{I}nönü-Wigner contraction procedure $\sl(2,\dsR) \simeq \mathfrak{so}(2,1)\;\stackrel{\Lambda\to 0}{\longrightarrow}\;\mathfrak{iso}(1,1)$ is performed by rescaling the transvection generators of $\mathfrak{so}(2,1)$, aka. the ladder generators of $\sl(2,\dsR)$, as $P_\pm\mapsto R \,P_\pm$, and taking the flat limit $R\to\infty$. The same contraction can be applied to the special linear algebras $\sl(N,\dsR)$, thought of as higher-spin algebras. Taking inspiration from what is done in three dimensions (see e.g. \cite{Afshar:2013vka,Grumiller:2014lna,Campoleoni:2015qrh,Campoleoni:2016vsh}) one may consider, for instance, the following prescription: rescale the ladder generators, $P^s_m := \varepsilon\, V^{s}_m$ for $m \neq 0$, while keeping unscaled the $N-1$ Cartan generators $V^{s}_0$, with $s=2,3,\ldots,N$; then taking the limit $\varepsilon \to 0$ defines the contracted algebra. In the contraction limit, one finds that the higher-spin translation generators $P^s_m$ form an Abelian ideal and have fixed Lorentz weight, $[J, P^s_m] = m\, P^s_m$. 

More generally speaking, the resulting algebra has the structure of a semi-direct sum, $\g=\mathfrak{h}\inplus\mathfrak{i}$, where $\mathfrak{h}$ is an Abelian subalgebra spanned by the $N-1$ higher-spin Lorentz generators $V_0^s$ and acting on the Abelian ideal $\mathfrak{i}$ spanned by the higher-spin translation generators $P^s_m$. This provides a finite-dimensional higher-spin generalisation of the two-dimensional Poincaré algebra $\mathfrak{iso}(1,1)$. Therefore, the BF action \eqref{eq:BF_action_pairing} defines a two-dimensional higher-spin version of dilaton gravity with vanishing cosmological constant. Its spectrum is a finite tower of topological gauge fields of spin ranging from 2 to $N$.
\minisec{Infinite-Dimensional Higher-Spin Algebras}
Another generalisation of BF theory to higher-spin gravity has been performed for the case $\Lambda<0$ in \cite{Alkalaev:2014qpa} and amounts to an introduction of an infinite number of topological gauge fields and dilaton-like fields, which can be achieved by constructing the higher-spin algebra\footnote{For the sake of simplicity, we are here not very precise in distinguishing (1) the associative algebra from the corresponding Lie algebra and (2) the latter algebra $\mathfrak{gl}[\lambda]$ from the higher-spin algebra $\hs[\lambda]$ resulting after quotienting the centre $\mathfrak{u}(1)$ spanned by the unit element; usually one writes $\mathfrak{gl}[\lambda]=\mathfrak{u}(1)\oplus\hs[\lambda]$. We refer to \cite{Alkalaev:2014qpa} for a careful presentation of these technical points in the context of two-dimensional higher-spin gravity.} as a quotient of the universal enveloping algebra of $\so(2,1)$ \cite{Feigin,Bordemann:1989zi,Bergshoeff:1989ns,Pope:1989sr},
\begin{align}\label{defhslambda}
    \hs[\lambda]=\frac{\mathcal{U}\big(\so(2,1)\big)}{\mathcal{C}-c\idty\simeq 0}\,, && \mathcal{C}=-\Lambda J^2+\frac{1}{2}\left\{P_+\,,P_-\right\}\,, && c\equiv \frac{\lambda^2-1}{4}\,,
\end{align}
where $\mathcal{C}$ denotes the quadratic Casimir element of $\mathcal{U}\left(\so(2,1)\right)$.\footnote{\label{noncomgeom} As an associative algebra, the quotient \eqref{defhslambda} has a natural interpretation in non-commutative geometry as the algebra of functions on the fuzzy hyperboloid (see e.g. \cite{Bergshoeff:1989ns,Vasiliev:1989re}).} In the seminal  construction \cite{Alkalaev:2014qpa}, the idea was to make use of an invariant metric (arising from a trace operation) on the higher-spin algebra (see \cite{Vasiliev:1989re,Alkalaev:2014qpa} for details) in order to generalise the BF action in its standard form \eqref{eq:BF_action_tr} in which both the one-form $A$ and the zero-form $B$ take values in $\hs[\lambda]$. Accordingly, the equations of motion remain of the same form $F=0$ and $\operatorname{D}\!B=0$. However, let us stress that an equivalent formulation in terms of the action \eqref{eq:BF_action_pairing}, i.e. involving the dual of $\hs[\lambda]$, is possible in the higher-spin case, as well.
\minisec{Extended Higher-Spin Algebras}
A further extension \cite{Alkalaev:2019xuv,Alkalaev:2020kut} of the theory allows to describe an infinite multiplet of massive local degrees of freedom. The starting point is an involutive automorphism $\tau$ of $\so(2,1)$ that can be explicitly defined through its action on the basis generators: $\tau(J)=J$ and $\tau(P_\pm)=-P_\pm$. This straightforwardly extends to the higher-spin algebra if the condition $\tau(\idty)=\idty$ is imposed and -- since the Casimir element $\mathcal{C}$ is left invariant by $\tau$ -- the automorphism descends to the quotient $\hs[\lambda]$, as well. Accordingly, the \emph{twisted-adjoint} action of the algebra on itself can be defined,
\begin{align}\label{eq:def_tad}
    \tad{_X}(Y):=X\cdot Y-Y\cdot\tau(X)\,.
\end{align}
Particularly, the twisted-adjoint action of the AdS-Lorentz transformation $X=J$ is identical to the adjoint one (i.e. the commutator), $\tad{_J}=\ad_J=[J\,,\,.\,]$, and that of transvections $X=P_\pm$ simply becomes the anti-commutator, $\tad{_{P_\pm}}\equiv\tad{_\pm}=\{P_\pm\,,\,.\,\}$.

The twist automorphism allows to define an \emph{extended} higher-spin algebra as the so-called smash product $\hs[\lambda]\rtimes\dsZ_2$, where $\dsZ_2=\{\idty,\tau\}$. Its elements are pairs $(a,b)$ with $a,b\in\hs[\lambda]$ that can be written $a\bldidty+b\bldtau$ and obey the multiplication rule
\begin{align}
    \left(a_1\bldidty+b_1\bldtau\right)\star\left(a_2\bldidty+b_2\bldtau\right)=\left(a_1\cdot a_2+b_1\cdot\tau(b_2)\right)\bldidty+\left(a_1\cdot b_2+b_1\cdot\tau(a_2)\right)\bldtau\,.
\end{align}
The adjoint action of the extended algebra on itself is defined as the commutator w.r.t. this product and will be denoted $\ead{_X}$; when restricted to $X\in\hs[\lambda]$ the adjoint action neatly decomposes as
\begin{align}
    \ead{_X}\left(a\bldidty+b\bldtau\right)=\ad_X(a)\bldidty+\tad{_X}(b)\bldtau\,;
\end{align}
i.e. it unifies the adjoint and twisted-adjoint action.

Now, at this point we would like to take the novel perspective of the dual-space formulation, which is why we have to introduce a few more concepts. The dual space to the extended higher-spin algebra can be defined in terms of a pairing that decomposes as
\begin{align}\label{eq:dual_pairing}
    \left(a^*\bldidty+b^*\bldtau\,,a\bldidty+b\bldtau\right)=\left(a^*,a\right)+\left(b^*,b\right)\,.
\end{align}
Furthermore, the respective coadjoint action $\ead{^*_X}$ is defined in the usual way,
\begin{align}
    \left(\ead{^*_X}(a^*\bldidty+b^*\bldtau)\,,a\bldidty+b\bldtau\right):=-\left(a^*\bldidty+b^*\bldtau\,,\ead{_X}(a\bldidty+b\bldtau)\right)\,.
\end{align}
If we moreover define the coadjoint analogue to the twisted-adjoint, namely
\begin{align}\label{eq:def_twisted-coadj}
    \left(\tad{^*_X}\left(Y^*\right)\,,Y\,\right):=-\left(Y^*,\tad{_X}(Y)\right)\,,
\end{align}
then the extended coadjoint action of $X\in\hs[\lambda]$ decomposes into the coadjoint and twisted-coadjoint action,
\begin{align}
    \ead{^*_X}\left(a^*\bldidty+b^*\bldtau\right)=\ad^*_X(a^*)\bldidty+\tad{^*_X}(b^*)\bldtau\,.
\end{align}
That is, the extended coadjoint action on $(\hs[\lambda]\rtimes\dsZ_2)^*$ unifies the coadjoint and twisted-coadjoint actions on $\hs[\lambda]^*$.

The extended action is of the form \eqref{eq:BF_action_pairing} equipped with the pairing \eqref{eq:dual_pairing} and is now a functional of the extended fields
\begin{align}
    \mathcal{B}^*=B^*\bldidty+C^*\bldtau\,, && \mathcal{A}=A\bldidty+Z\bldtau\,.
\end{align}
The extended fields satisfy $\d\mathcal{A}+\mathcal{A}\wedge_{\!\star}\mathcal{A}=0$ and $\d\mathcal{B}^*+\ead{^*_{\mathcal{A}}}(\mathcal{B}^*)=0$, the former decomposing into
\begin{subequations}\label{eqs:eom_A_Z}
\begin{align}
    \d A+A\wedge A+Z\wedge\tau(Z)&=0\,,\\
    \d Z+A\wedge Z+Z\wedge\tau(A)&=0\,.
\end{align}
\end{subequations}
From these equations it can be argued that the one-form $Z$ can be set to zero on-shell by means of a gauge transform, leading back to the one-form sector obeying gauge flatness, i.e. $\d A+A\wedge A=0$, which describes solutions that are locally (higher-spin) AdS$_2$. In this gauge the components of the zero-forms are covariantly constant w.r.t. the coadjoint and twisted-coadjoint action, respectively:
\begin{align}\label{eqs:covariant_constancy}
    \d B^*+\ad^*_A\left(B^*\right)=0\,, && \d C^*+\tad{^*_A}\left(C^*\right)=0\,.
\end{align}
The latter equation is the unfolded description of an infinite tower of scalar fields of ever increasing mass, as we will argue in the following subsection. It does however appear easier to translate the second equation in \eqref{eqs:covariant_constancy} from the dual algebra to the algebra itself by introducing another zero-form $C\in\hs[\lambda]$ in terms of an invertible mapping $\Phi:\,\hs[\lambda]\rightarrow\hs[\lambda]^*$ that intertwines the adjoint and coadjoint representation (as well as the twisted-adjoint and twisted-coadjoint representation) and the existence of which is ensured in the semisimple case. Then defining $C=\Phi^{-1}\cdot C^*$ and fixing its norm through $(C^*,C)=\langle C\,,C\rangle=1$, the matter field in $\hs[\lambda]$ fulfils the well-known unfolded equation
\begin{align}\label{eq:eom_C}
    \d C+\tad{_A}(C)=0\,.
\end{align}
\minisec{Klein-Gordon Equation}
At this point it is worthwhile having a look at equation \eqref{eq:eom_C} and how it encodes scalar degrees of freedom on a classical background from a rather general perspective: For any value of the cosmological constant (and for both the two- and three-dimensional case) the d'Alembert operator, when acting on a scalar, can be expanded in the frame-like basis as
\begin{align}
    \Box=\eta^{ab}e^\mu_a e^\nu_b \partial_\mu\partial_\nu+\eta^{ab}e^\mu_a\left(\partial_\mu e^\nu_b\right)\partial_\nu-e^\mu_ae^\nu_b\varvarpi_\mu^{ab}\partial_\nu\,.
\end{align}
Here, vielbein and spin connection are defined in the usual way, i.e.
\begin{align}
    g_{\mu\nu}=e^a_\mu e^b_\nu \eta_{ab}\,, && \varvarpi_\mu{}^a{}_b=e^a_\rho e^\nu_b\Gamma^\rho_{\mu\nu}-e^\nu_b\left(\partial_\mu e_\nu^a\right)\,,
\end{align}
and, depending on the dimension of the spacetime,
\begin{align}
\varvarpi_\mu^{ab}=\begin{cases}
        \varepsilon^{ab}\varvarpi_\mu\,,           & d=2\,,\\
        \varepsilon^{ab}{}_c\varvarpi^c_\mu\,,     & d=3\,.
    \end{cases}
\end{align}
When the gauge field $A$ describes a classical (spin-two, i.e. JT- or Einstein gravity) background, it takes values in $\so(d,1)$ or $\iso(d-1,1)$ and can be decomposed into the algebra-valued one-forms $\omega=d\!x^\mu\varvarpi_\mu^{ab}J_{ab}$ (only containing (AdS-)Lorentz transformations) and $e=d\!x^\mu e_\mu^a P_a$ (only containing transvections/translations), such that $A=\omega+e$, and equation \eqref{eq:eom_C} can be written in one-form  components as
\begin{align}
    \partial_\mu C+\left[\omega_\mu\,,C\right]+\left\{e_\mu\,,C\right\}=0\,.
\end{align}
One may then consider the action of the scalar d'Alembert operator on the matter field, which results in
\begin{align}\label{eq:unfolded_box}
    \Box C=\eta^{ab}\big\{P_a\,,\left\{P_b\,,C\right\}\big\}+\dots\,,
\end{align}
where the dots denote terms that contain commutators with the spin connection $\omega_\mu$ or its derivatives, only. As we are going to discuss in the following subsection for the case $d=2$, there exists a decomposition of the respective higher-spin algebra such that certain components of $C$ decouple from the rest, namely they do not appear in ``$\dots$'', while the nested anti-commutator gives rise to a mass term.
\subsection{The Twisted-Adjoint Basis and the Mass Spectrum}\label{subsec:tw_basis_spectrum}
When the background geometry is restricted to spin two, i.e. $A\in\so(2,1)$, we can study the field content of equation \eqref{eq:eom_C} by analysing the module structure of $\hs[\lambda]$ under the twisted-adjoint action of $\so(2,1)$. This has been studied in \cite{Alkalaev:2019xuv}, but we are going to follow a different approach here, one that builds upon the explicit construction of a basis. Throughout this subsection we keep the cosmological constant $\Lambda$ unspecified to better understand in which sense $\Lambda=0$ will turn out to be peculiar.
\minisec{The Adjoint Basis}
Let us first sketch how to construct a basis choice that reflects the module structure of $\hs[\lambda]$ under the adjoint action. First, any basis element can be assumed to consist of a fixed number $m$ of identical transvections -- since the equivalence relation parametrising the Casimir element is imposed, cf. the second relation in \eqref{defhslambda}, we can think of mixed expressions such as $P_+P_-$ as being eliminated in favour of powers of $J$. Thus one considers basis elements proportional to $(P_\pm)^m$ times a polynomial in $J$ and, therefore, the adjoint action of $J$ on such a generator gives the eigenvalue $\pm m$. (Accordingly, such a generator will sometimes be sloppily called a ``$\pm m$ mode''.) Then, a second index $s$ shall be introduced in order to, first, distinguish linearly independent combinations of powers of $J$ and, second, label different sub-modules; the latter means that $s$ is not allowed to change under the adjoint action of any element of $\so(2,1)$. Then a reasonable ansatz is to define
\begin{align}\label{eq:def_adj_generators}
    V^s_{\pm m}:=\ad^m_\pm\left(V^s_0\right)\,, && V^s_0=V^s_0(J)\,,
\end{align}
where $V^s_0$ denotes functions of $J$ that are linearly independent for different $s$ (i.e. those generators form a basis of $\mathcal{U}\big(\so(1,1)\big)$. The Lorentz generator automatically acts as $[J\,,V^s_{\pm m}]=\pm m V^s_{\pm m}$. Furthermore, by construction the transvection generators act as $[P_\pm\,,V^s_{\pm m}]=V^s_{\pm (m+1)}$. The remaining condition that we impose on the basis is the closure under the mixed-sign combination for a given $s$: 
\begin{align}\label{eq:inv_cond_ad_hs}
    \left[P^{}_\mp\,,V^s_{\pm m}\right]\stackrel{!}{=}a_m(s)V^s_{\pm (m-1)}\,, && m\geqslant 1\,.
\end{align}
The base case is $m=1$, then the remaining conditions for $m>1$ follow by induction, namely the case $m+1$ is traced back to its predecessor by using the definition of $V^s_{\pm m}$ and Jacobi's identity:
\begin{align}
    \begin{split}
    \Big[P_{\mp}\,,V^s_{\pm(m+1)}\Big]&=\Big[P_\mp\,,\big[P_\pm\,,V^s_{\pm m}\big]\Big]=\Big[P_\pm\,,\big[P_\mp\,,V^s_{\pm m}\big]\Big]+\Big[\big[P_\mp\,,P_\pm\big]\,,V^s_{\pm m}\Big]\\
    &=a_m(s)\left[P_{\pm}\,,V^s_{\pm(m-1)}\right]\pm 2\Lambda \left[J\,,V^s_{\pm m}\right]\\
    &=\big(a_m(s)+2\Lambda m\big) V^s_{\pm m}\,,
    \end{split}
\end{align}
where we used the assumption \eqref{eq:inv_cond_ad_hs}; as a side result we find $a_m(s)=\mu(s)+\Lambda m(m-1)$, where the initial value is abbreviated $a_1(s)\equiv\mu(s)$. The invariance condition \eqref{eq:inv_cond_ad_hs} for $m=1$ leads to the condition 
\begin{align}\label{eq:algebraic_equation_hs_adj}
    2\big(c+\Lambda J^2\big)V^s_0(J)-\big(c+\Lambda J(J+1)\big)V^s_0(J+1)-\big(c+\Lambda J(J-1)\big)V^s_0(J-1)=\mu(s)V^s_0(J)\,,
\end{align}
with the proportionality constant $\mu(s)$ as introduced above. This equation allows for solutions that are polynomials of order $s-1$ in $J$, which can be seen directly by inspection of \eqref{eq:algebraic_equation_hs_adj}. A polynomial of order $s-1$ provides a maximum of $s$ equations to fix the respective coefficients but the overall normalisation can be freely chosen; thus, ideally there is one equation left to fix the coefficient $\mu(s)$, as well. By direct computation, one can check that, upon setting $\Lambda=-1$ and $c=(\lambda^2-1)/4$, the zero-mode generator\footnote{The explicit form, up to normalisation, reads (see Footnote~\ref{fn:factorials} for the explanation on the notation):
\begin{align}
    V^s_0(J)\sim\sum_{n=0}^{s-1}\frac{(-1)^n}{n!^2}\frac{(s+n-1)!}{(s-n-1)!}\frac1{(1-\lambda)_n}\left(J+\frac{1-\lambda}{2}\right)_n\,.
\end{align}
}%
 of the commonly used \emph{highest-weight basis} \cite{Fradkin:1990ir} of $\hs[\lambda]$ solves equation \eqref{eq:algebraic_equation_hs_adj} with $\mu(s)=s(s-1)$.

This basis choice makes clear that, as a representation space, the higher-spin algebra is highly reducible and decomposes into an infinite, discrete collection of finite-dimensional $\so(2,1)$-modules, each of which consists of finitely-many one-dimensional $\so(1,1)$-modules:
\begin{align}
    \hs[\lambda]=\bigoplus_{s=1}^\infty\mathcal{V}^{(s)}\,, && \mathcal{V}^{(s)}=\bigoplus_{m=-(s-1)}^{s-1}\mathcal{V}^{(s)}_m\,,
\end{align}
where $\mathcal{V}^{(s)}\simeq\mathcal{D}_{s-1}$ are finite-dimensional irreducible $\so(2,1)$-modules ($\dim \mathcal{V}^{(s)}=2s-1$), while the one-dimensional $J$-eigenspace $\mathcal{V}^{(s)}_m$ is spanned by $V^s_m$.
\minisec{The Twisted-Adjoint Basis}
One may try and mimic the previous construction of a basis adapted to the twisted-adjoint action. We therefore define
\begin{align}\label{eq:def_tadj_generators}
    W^k_{\pm m}:=\tad{^m_\pm}\left(W^k_0\right)\,, && W^k_0=W^k_0(J)\,,
\end{align}
where the generators $W^k_0$ are once more supposed to form a basis of $\mathcal{U}\big(\so(1,1)\big)$. Note the important difference to the adjoint basis \eqref{eq:def_adj_generators}: we will see that powers of the twisted-adjoint action do not truncate for a given $k$, in other words the index $m$ can take unbounded integer values, regardless of $k$. By construction, $\tad{_\pm}(W^k_{\pm m})=\{P_\pm\,,W^k_{\pm m}\}=W^k_{\pm(m+1)}$ hence, again, since the number of positive- or negative-index transvections is fixed, the twisted-adjoint action of $J$ -- which agrees with its adjoint action -- is simply $\tad{_J}(W^k_{\pm m})=[J,W^k_{\pm m}]=\pm m W^k_{\pm m}$. The remaining invariance condition of the module spanned by elements with fixed index $k$ reads
\begin{align}\label{eq:inv_cond_tad_hs}
    \left\{P^{}_\mp\,,W^k_{\pm m}\right\}\stackrel{!}{\sim}W^k_{\pm (m-1)}\,, && m\geqslant 1\,,
\end{align}
which again can be obtained by induction\footnote{The generalised Jacobi identity $\{X\,,\{Y\,,Z\}\}=\{Y\,,\{Z\,,X\}\}-[Z\,,[X\,,Y]]$ proves to be useful.} from the base case $m=1$; the latter leads to
\begin{align}\label{eq:algebraic_equation_hs_tadj}
    2\big(c+\Lambda J^2\big)W^k_0(J)+\big(c+\Lambda J(J+1)\big)W^k_0(J+1)+\big(c+\Lambda J(J-1)\big)W^k_0(J-1)=M^2(k)W^k_0(J)\,,
\end{align}
where now the constant of proportionality -- with hindsight -- is called $M^2(k)$. Note that the difference to equation \eqref{eq:algebraic_equation_hs_adj} is merely a sign in the last two summands on the left-hand side but it changes the solution space drastically. In particular, equation \eqref{eq:algebraic_equation_hs_tadj} does not possess polynomial solutions, which is in accordance with the findings of \cite{Alkalaev:2019xuv}.

Although we are at present not able to display a closed-form solution for $W^k_0(J)$ in full generality, it is possible to take the ansatz\footnote{\label{fn:factorials}The symbol $(x)_n=(x+n-1)\cdots(x+1)x=\Gamma(x+n)/\Gamma(x)$ denotes the Pochhammer symbol (aka. rising factorial). To derive \eqref{recurrencerel}, one basically uses identities such as $(x+n)\,(x)_n=(x)_{n+1}$ and $(x-1)_{n+1}=(x)_n\,(x-1)\,$.}
\begin{align}
    W^k_0(J)=\sum_{n=0}^\infty \frac{\alpha_n}{n!^2}(J+\Delta)_n\,, && \Delta\equiv \frac{\lambda+1}{2}
\end{align}
for $\Lambda=-1$ and derive a finite-term recurrence relation for $\alpha_n$. Explicitly:
\begin{align}\label{recurrencerel}
\begin{split}
    & (2\Delta+n)\alpha_{n+1}-\Big(2(2\Delta+n)(2n+1)+n(n-1)
    +M^2\Big)\alpha_n\\
    &\quad\quad\quad\quad\quad +4n^2(2\Delta+2n-1)\alpha_{n-1}-4n^2(n-1)^2
    \alpha_{n-2}=0\,.
\end{split}
\end{align}
There may exist several classes of solutions and by choosing one of these classes one effectively defines a  completion of the algebra (as a topological vector space).

The existence of the twisted-adjoint basis makes transparent how (a completion of) the higher-spin algebra decomposes into irreducible, infinite-dimensional $\so(2,1)$-modules $\mathcal{W}^{(k)}$ in the twisted-adjoint representation (the so-called \emph{Weyl modules}), each of which decomposing into one-dimensional $\so(1,1)$-modules,
\begin{align}
    \widehat{\hs}[\lambda]=\bigoplus_{k=0}^\infty \mathcal{W}^{(k)}\,, && \mathcal{W}^{(k)}=\bigoplus_{m\in\dsZ}\mathcal{W}^{(k)}_m\,.
\end{align}
The one-dimensional $J$-eigenspace $\mathcal{W}^{(k)}_m$ is spanned by the generator $W^k_m$.
\minisec{Mass Spectrum}
We can now come back to the unfolded equation and its implication \eqref{eq:unfolded_box}. Being interested in terms proportional to the zero-mode generators $W^k_0$, it is easy to see that the suppressed terms in that equation -- those which contain commutators with (derivatives of) the spin connection -- cannot produce any such generator because the sum of modes is conserved under commutation and the eigenvalue of the adjoint action of $J$ on $W^k_0$ is zero. Expanding the matter field as
\begin{align}
    C=\sum_k\left(\phi^{(k)}W^k_0+\sum_\pm\sum_{m\in\dsN}c^{k,m}_\pm W^k_{\pm m}\right)\,,
\end{align}
the action of the d'Alembert operator on zero-mode coefficients reads
\begin{align}
    \Box\phi^{(k)}=\text{coefficient of }W^k_0\text{ in}\left(\frac{1}{2}\big\{P_+\,,\left\{P_-\,,C\right\}\big\}+\frac{1}{2}\big\{P_-\,,\left\{P_+\,,C\right\}\big\}\right)\,.
\end{align}
From the defining property \eqref{eq:algebraic_equation_hs_tadj} it follows that
\begin{align}
    \left(\Box-M^2(k)\right)\phi^{(k)}=0\,.
\end{align}
This shows that equation \eqref{eq:eom_C} describes an infinite tower of scalar fields of different masses; although the explicit form of the mass spectrum could not be extracted in the construction above (this would require a general classification of solutions of \eqref{eq:algebraic_equation_hs_tadj}, which we have not performed), the results of \cite{Alkalaev:2019xuv} for $\Lambda=-1$ provide us with the relation 
\begin{equation}\label{discretespectrum}
    M^2(k)=\frac{(k-\lambda)(k-\lambda+1)}{R^2}\quad\text{for}\quad k=0,1,2,\dots\,.
\end{equation}
\section{Vanishing-Cosmological-Constant Higher-Spin Gravity Coupled to Matter}\label{sec:flat_case}
Within this section, we transfer all previous considerations to the case of vanishing cosmological constant. Many of the ingredients discussed above are straightforwardly applicable to this case, but we are going to repeat the most important points.
\subsection{Higher-Spin Algebra and Twisted-Adjoint Basis}\label{subsec:ihs_tw_basis}
As in the case of AdS$_2$, the starting point for the construction of a higher-spin algebra is the spin-two isometry algebra, here $\iso(1,1)$. As a basis we use the Lorentz boost $J$ and translations $P_\pm$, equipped with the Lie brackets
\begin{align}\label{eq:Lie_bracket_Poincare}
    \left[J\,,P_\pm\right]=\pm P_\pm\,, && \left[P_\pm\,,P^{}_\mp\right]=0\,,
\end{align}
which has already been contained as the case $\Lambda=0$ in \eqref{eq:algebra_generalL}. Obviously, translations form a non-trivial, Abelian ideal $\dsR^2\subset\iso(1,1)$ and the Poincaré algebra is not semisimple.

The second-order Casimir element within the universal enveloping algebra of $\iso(1,1)$ is the mass squared $\mathcal{M}^2=P_+P_-$, which can be set to a multiple of the identity,\footnote{If the Poincaré algebra is viewed as the \.{I}nönü-Wigner contraction $\so(2,1)\xrightarrow{\Lambda\to 0}\iso(1,1)$, then one can write the effective $\so(2,1)$-Casimir as $\mathcal{C}=(\scrim^2/4)\idty$. Note that the dimensionless quantity that tends to zero is $\varepsilon\equiv\Lambda/\scrim^2$. From $\scrim^2=-\Lambda(\lambda^2-1)$ it follows that $|\lambda|$ needs to run to infinity. Therefore, there is the following \.{I}nönü-Wigner contraction:
\begin{align}
    \hs[\lambda]\;\xrightarrow[|\Lambda|\lambda^2\,\sim\,\scrim^2>0]{|\lambda|\to\infty,\;\Lambda\to 0}\; \ihs[\scrim]\,.
\end{align}
} thereby defining the quotient
\begin{align}\label{ihsM}
    \ihs\left[\scrim\right]=\frac{\mathcal{U}\big(\iso(1,1)\big)}{\mathcal{M}^2-c\idty\simeq 0}\,, && c=\frac{\scrim^2}{4}\,.
\end{align}
Note that the above quotients are isomorphic $\ihs\left[\scrim_1\right]\simeq\ihs\left[\scrim_2\right]$ for any $\scrim_1>0$ and $\scrim_2> 0$, since one can rescale $P_\pm$ without affecting the commutation relations \eqref{eq:Lie_bracket_Poincare}. This feature must be contrasted with the one-parameter family of inequivalent higher-spin algebras $\hs[\lambda]$ (for which $\hs[\lambda_1]\simeq\hs[\lambda_2]$ iff $\lambda_1+\lambda_2= 1$).
We nevertheless keep the notation $\ihs[\scrim]$ to emphasise the analogy with $\hs[\lambda]$.

On this algebra $\ihs[\scrim]$, we define exactly the same structures as before, namely it carries an adjoint and a twisted-adjoint action, both of $\ihs[\scrim]$ itself and of its subalgebra $\iso(1,1)$. This turns $\ihs[\scrim]$ into an infinite-dimensional, highly reducible module for the respective representations of $\iso(1,1)$. In the following, we will discuss possible decompositions of $\ihs[\scrim]$ into $\iso(1,1)$-submodules by means of constructing suitable bases.
\minisec{The Adjoint Basis}
We take the same definition \eqref{eq:def_adj_generators} as in the case of $\hs[\lambda]$. Then the invariance condition is again \eqref{eq:inv_cond_ad_hs} and it again leads to \eqref{eq:algebraic_equation_hs_adj} but now in the case $\Lambda=0$, which changes its properties drastically. Here, inserting $c=\scrim^2/4$:
\begin{align}\label{eq:algebraic_equation_ihs_adj}
    2V^s_0(J)-V^s_0(J+1)-V^s_0(J-1)=\frac{4\mu(s)}{M^2}V^s_0(J)\,.
\end{align}
The only polynomial solutions to this equation are affine functions $V^s_0(J)=\alpha J+\beta\idty$ with $\mu(s)=0$, meaning that there exists no highest-weight basis in the usual sense. This can be seen by contradiction: Assuming $V^s_0(J)$ to contain a contribution $a_n J^n$ at highest power $n$, then at leading order one finds $\mu=0$. At next-to-leading order there is no condition, while at next-to-next-to leading order one finds $n(n-1)a_n=0$, allowing for non-trivial solutions only at orders $n=0$ and $n=1$, i.e. affine functions of $J$. This would lead to the basis element $\idty$ of the trivial representation, and to the basis elements $J$ and $P_\pm$ of the adjoint representation of $\mathfrak{iso}(1,1)$ on itself. 

However, there do exist power-series solutions of \eqref{eq:algebraic_equation_ihs_adj}, one example being
\begin{align}
    V^s_0(J)=\e{-s J}\,, && \mu(s)=-\scrim^2\sinh^2\left(\frac{s}{2}\right)\,.
\end{align}
Formally, both from the linear and exponential solutions one can construct an additional class carrying periodic or anti-periodic factors $\exp(\i\pi n J)$, $n\in\dsZ$, which modify $\mu(s)$ if $n$ is an odd number.  Given the discussions presented in Appendix~\ref{app:sec:systematic_app}, we can be reasonably confident that those cases do indeed cover all possible types of solutions of equation \eqref{eq:algebraic_equation_ihs_adj}. If the solutions are required to form a basis of a suitable completion of the universal enveloping algebra $\mathcal{U}\big(\so(1,1)\big)$, then the linear solutions are obviously not sufficient. Moreover, purely periodic solutions have to be discarded because they lead\footnote{In fact, for solutions such that $V^s_0(J)=V^s_0(J\pm 1)$, commuting a translation generator $P_\pm$ past any basis element $V^s_0(J)$ will not affect it and, thus, the adjoint action of $\iso(1,1)$ is trivial on the span of these elements.} to a trivial adjoint action of translations, so they do not help to construct the basis elements $V^s_{\pm m}(J)$ with $m\neq 0$.

In any case, there exists a decomposition of a suitable completion of the flat-space higher-spin algebra into subspaces $\mathcal{V}^{(s)}$ that are invariant under the adjoint action of $\iso(1,1)$ and irreducible by construction; they decompose into one-dimensional $\so(1,1)$-modules,
\begin{align}
    \widehat{\ihs}[\scrim]=\bigoplus_{s\in  [0,\infty)} \mathcal{V}^{(s)}\,, && \mathcal{V}^{(s)}=\bigoplus_{m\in\dsZ}\mathcal{V}^{(s)}_m\,.
\end{align}
For any positive number $s>0$, the adjoint module $\mathcal{V}^{(s)}$ growing from non-polynomial solution $V^s_0(J)=\e{-s J}$ is spanned by the basis elements $V^s_{\pm m}$ with unbounded values of $m$. These adjoint modules carry infinite-dimensional irreducible representations. This is a feature, not a bug. 

In fact, let us briefly review a few basis facts about irreducible representations of $\iso(1,1)$. First, there are \emph{no} highest-weight (or lowest-weight) representations of $\iso(1,1)$ that are both irreducible and have a non-trivial action of the translations. This holds because $\ker P_+$ (or $\ker P_-$) is an invariant subspace, so it is either zero or the whole module. Second, any finite-dimensional irreducible representation of  $\iso(1,1)$ is one-dimensional and unfaithful, with trivial action of translations and diagonal action of boosts. This follows from Schur's lemma and the fact that any finite-dimensional representation must be highest (and lowest) weight. Third, any faithful finite-dimensional representation of $\mathfrak{iso}(1,1)$ is indecomposable (in the sense of reducible but not fully reducible). This is for instance the case for Killing-tensor modules: they are finite-dimensional and indecomposable. The strategy for constructing the adjoint basis for any $\Lambda$ is to build irreducible modules $\mathcal{V}^{(s)}$ growing from a zero-mode by repeated action of translation generators. From the previous facts, it becomes clear that such irreducible modules $\mathcal{V}^{(s)}$ are necessarily infinite-dimensional, in contrast with the Killing-tensor modules. This does not prevent the existence of a \emph{Killing basis} where the basis elements are Weyl-ordered (i.e. totally-symmetrised) polynomials of degree $s-1$ in the generators and spanning an $\mathfrak{iso}(1,1)$-module $\mathcal{U}^{(s)}$ isomorphic to the space of Killing tensors of rank $r\leqslant s-1$. For instance, the set $\{1,P_+,P_-,J,P_+^2,P_+(J+\sfrac12),J^2,P_-(J-\sfrac12), P_-^2\}$  provides a basis of $\mathcal{U}^{(3)}$. There is a filtration $\ihs[\scrim]=\bigcup_{s=1}^\infty \mathcal{U}^{(s)}$ of the higher-spin algebra with $\mathcal{U}^{(0)}\subset\mathcal{U}^{(1)}\subset\mathcal{U}^{(2)}\subset\ldots$, where each quotient $\mathcal{K}^{(s)}:=\mathcal{U}^{(s)}/\mathcal{U}^{(s-1)}$ is isomorphic to the space of Killing tensors of rank $s-1$ ($\dim\mathcal{K}^{(s)}=2s-1$). This remark provides a decomposition of the higher-spin algebra as follows:
\begin{align}
    \ihs[\scrim]\simeq\bigoplus_{s=1}^\infty \mathcal{K}^{(s)}\,, && \mathcal{K}^{(s)}=\bigoplus_{m=1-s}^{s-1}\mathcal{K}^{(s)}_m\,.
\end{align}
From the perspective of our basis definition, each $\mathcal{K}^{(s)}$ can still be thought of as being spanned by generators defined in terms of powers of the adjoint action \eqref{eq:def_adj_generators}, but we do not impose the invariance condition \eqref{eq:algebraic_equation_ihs_adj}, such that the zero-mode generators $V^s_0(J)$ can be taken to be polynomial functions of degree $s-1$ in $J$.

Before moving on, let us comment on possible truncations of the algebra. In contrast to $\hs[\lambda]$, which reduces (upon quotienting by an infinite-dimensional ideal) to $\sl(N,\dsR)$ when the parameter takes integer values, $\lambda=N$, there is no analogous statement in the flat-space case. Only for vanishing Casimir, $\scrim=0$, there appears an infinite tower of non-trivial ideals of the Lie algebra $\ihs[0]$, namely
\begin{align}
    \mathcal{I}^{(\pm)}_n=\operatorname{span}\left\{V^s_{\pm m}\,\middle|\,s\in\dsR\,,\ m\geqslant n\right\}\,.
\end{align}
This is due to the fact that the absolute value of the index $m$ cannot be lowered any more, since the respective commutation relation contains a factor of $\scrim$ through the identification of the Casimir element. These ideals are obviously all contained in each other, $\mathcal{I}^{(\pm)}_0 \supset \mathcal{I}^{(\pm)}_1 \supset \mathcal{I}^{(\pm)}_2 \supset\dots$, and the quotient w.r.t. the largest possible ideal, $\mathcal{I}^{(\pm)}_0$, would result in the trivial algebra.
\minisec{The Twisted-Adjoint Basis}
Now we want to study the analogous decomposition of $\ihs[\scrim]$ under the twisted-adjoint action; an obvious ansatz for that is to again use the definition \eqref{eq:def_tadj_generators}, from which follows the invariance condition
\begin{align}\label{eq:algebraic_equation_ihs_tadj}
    2W^k_0(J)+W^k_0(J+1)+W^k_0(J-1)=\frac{4M^2(k)}{M^2}W^k_0(J)\,.
\end{align}
Similarly to the adjoint case, the only solutions of polynomial type are affine functions $W^k_0(J)=\alpha J+\beta\idty$  with $M^2(k)=M^2$. As an example of a non-polynomial class of solutions there is again the set of exponential functions:
\begin{align}\label{eqs:exponential_basis}
    W^k_0(J)=\e{-k J}\,, && M^2(k)=M^2\cosh^2\left(\frac{k}{2}\right)\,.
\end{align}
Once more it is possible to add periodic or anti-periodic factors $\exp(\i\pi n J)$ with $n\in\dsZ$ to these solutions and the discussions of Appendix~\ref{app:sec:systematic_app} suggest that all solutions of \eqref{eq:algebraic_equation_ihs_tadj} are covered. Note however that, as soon as we allow a certain class of non-polynomial solutions, we are essentially fixing a particular completion of the algebra as a vector space and it is not identical with the original quotient of the universal enveloping algebra, any more. Furthermore note that for purely anti-periodic solutions, i.e. $W^k_0(J)=-W^k_0(J\pm 1)$, the anti-commutator with translations vanishes, implying $W^k_{\pm m}=0$ for $m\neq 0$. Accordingly, both periodic and antiperiodic classes of solutions have to be discarded if we want a completion defined by the same choice of basis, since these classes result in trivial representations of $\iso(1,1)$ for the adjoint and twisted-adjoint actions, respectively.

Concerning the possible values of $k$ it is now reasonable to consider $k\in [0,\infty)$, which means we are working with a continuous ``basis'' and it spans the set of functions that can be represented in terms of the (unilateral) Laplace transform w.r.t. $J$,
\begin{align}\label{eq:Laplace_decomp}
    f(J)=\int\limits_0^\infty\!\!\d k\,\tilde f(k)\e{-k J}\,.
\end{align}
Apparently, if we want to recover $\iso(1,1)$ itself, we have to allow for distributional coefficients: while translations are simply $W^0_\pm=2P_\pm$, in order to write $J$ we have to take $\tilde f(k)=\delta'(k)$ in \eqref{eq:Laplace_decomp}, i.e. we are instructed to work within the realm of \emph{generalised functions}. Through higher derivatives of the delta distribution we can make contact with the monomial basis.\footnote{This behaviour is not at all unusual in physics; the analogous problem arises in the passage from the overcomplete basis of canonical coherent states (of the quantum-harmonic oscillator, say) back to a discrete basis.}

The insight here is that there exists a decomposition of a suitable completion of the flat-space higher-spin algebra into subspaces $\mathcal{W}^{(k)}$ that are invariant under the twisted-adjoint action of $\iso(1,1)$ and irreducible by construction; they decompose into one-dimensional $\so(1,1)$-modules,
\begin{align}
    \widehat{\ihs}[\scrim]=\bigoplus_{k\in  [0,\infty)} \mathcal{W}^{(k)}\,, && \mathcal{W}^{(k)}=\bigoplus_{m\in\dsZ}\mathcal{W}^{(k)}_m\,.
\end{align}
The twisted-adjoint basis $W^k_{m}$ with $m\in\dsZ$ can be viewed as the Lorentzian analogue of the \emph{angular-momentum basis} of $\iso(2)$ considered in Chapter~9.2 of \cite{Tung}.

Finally, given the pivotal role played by the dual algebra, it is important to note that the definition of the completion $\widehat{\ihs}[\scrim]$ suggests to define a respective completed dual algebra $\widehat{\ihs}[\scrim]^*$ as the span of the dual basis. To be concrete, the latter is defined as the span of basis elements $W^{*k}_{\pm m}$, in turn defined implicitly via the pairing, $(W^{*k}_{m}\,,W^{k'}_{m'}):=\delta^{kk'}\delta_{m+m',0}$. It is then straightforward to identify the twisted-coadjoint action of $\iso(1,1)$ on $\widehat{\ihs}[\scrim]^*$, which reads
\begin{subequations}
\begin{align}
    \ad^*_J\left(W^{*k}_{\pm m}\right)&=\pm m W^{*k}_{\pm m}\,,\\
    \tad{^*_\pm}\left(W^{*k}_{\pm m}\right)&=-M(k)^2 W^{*k}_{\pm(m+1)}\,,\\
    \tad{^*_\mp}\left(W^{*k}_{\pm m}\right)&=-W^{*k}_{\pm(m-1)}\,.
\end{align}
\end{subequations}

In the sequel, we will sometimes be sloppy with notation and will not write explicitly the hat on the higher-spin algebra for the sake of simplicity, the suitable completion being implicitly understood (except when the distinction is important).
\subsection{BF-Theory and Unfolded Scalar Fields}\label{subsec:flat_BF_theory}
We have now collected all necessary pieces to write down a BF theory of an infinite collection of higher-spin gauge fields at vanishing cosmological constant, as well as its version extended by an infinite collection of scalar matter fields. 

The action \eqref{eq:BF_action_pairing}, where $A\in\ihs[\scrim]$ and $B^*\in\ihs[\scrim]^*$,  describes a BF-like higher-spin theory with the same equations of motion as before, namely gauge flatness as well as \eqref{eq:eom_dualB}. One may linearise the theory around an asymptotically flat vacuum solution $A_0$ \cite{Afshar:2019axx,Afshar:2021qvi}, that is $A=A_0+A_1$. Then at linear order in $A_1$ one finds
\begin{align}
    \d A_1+A_0\wedge A_1+A_1\wedge A_0=0\,,
\end{align}
which is just the vanishing of the standard covariant derivative w.r.t. $A_0$, acting on $A_1$. Since the vacuum solution for the zero-form is $B^*=0$, the term $\ad^*_{A_1}(B^*)$ is of higher than linear order and, thus, the dual version of the standard covariant derivative of the generalised dilaton field vanishes, as well:
\begin{align}\label{eq:Bstar_coadjoint}
    \d\!B^*+\ad^*_{A_0}\left(B^*\right)=0\,.
\end{align}
Therefore the degrees of freedom contained in $B^*$, i.e. the higher-spin generalisations of the dilaton field of JT gravity, are \emph{global} degrees of freedom.

To analyse equation \eqref{eq:Bstar_coadjoint}, it is convenient to use the adjoint basis introduced above. We have seen that (the completion of) the higher-spin algebra decomposes into \emph{infinite}-di\-men\-sio\-nal modules $\mathcal{V}^{(s)}$ under the adjoint action of $\iso(1,1)$. This is a consequence of equation \eqref{eq:algebraic_equation_ihs_adj} not allowing for a polynomial basis, and this property translates to the coadjoint representation (on a dual space defined w.r.t. to the completed vector space), as well.  Accordingly, there may appear to be a mismatch of the number of degrees of freedom in equation \eqref{eq:Bstar_coadjoint} since for the adjoint representation of the higher-spin algebra one would have expected to find dilatonic fields in one-to-one correspondence with Killing tensor fields in flat spacetime, spanning $\iso(1,1)$-modules $\mathcal{K}^{(s)}$ of dimension $2s-1$, as would follow from the $\Lambda\to0$ limit of \cite{Alkalaev:2014qpa, Alkalaev:2020kut}. This apparent mismatch is just the price we pay for performing a decomposition of $\widehat{\ihs}[\scrim]$ in terms of the infinite-dimensional irreducible modules $\mathcal{V}^{(s)}$ rather than the finite-dimensional spaces $\mathcal{K}^{(s)}$. However, in the original algebra ${\ihs}[\scrim]$ nothing prevents us from using the latter decomposition instead, thereby resolving the tension: being in possession of a filtration of the algebra is sufficient to extract the right number of degrees of freedom from \eqref{eq:Bstar_coadjoint}.

Let us now have a look at the matter sector. In the extended version of the theory  $\g=\ihs[\scrim]\rtimes\dsZ_2$ and the action \eqref{eq:BF_action_pairing} as well as the equations of motion \eqref{eqs:eom_A_Z} are still of the same form, eventually leading (in the gauge choice $Z=0$) to the covariant-constancy condition \eqref{eqs:covariant_constancy}. That is, we immediately see that the theory contains a collection of unfolded massive, scalar fields. Using the decomposition of the higher-spin algebra into $\iso(1,1)$-submodules under the twisted-adjoint action, i.e. the twisted basis implicitly defined in \eqref{eq:algebraic_equation_ihs_tadj}, the zero-mode coefficients $\phi^{(k)}$ in the matter field fulfil
\begin{align}
    \left(\Box-M^2(k)\right)\phi^{(k)}=0\,.
\end{align}
Concerning the mass spectrum $M(k)$ we want to discuss two examples.
\begin{itemize}
    \item One may choose the exponential basis \eqref{eqs:exponential_basis}, which leads to an infinite set of scalar fields of ascending mass $M^2(k)=M^2\cosh^2(\sfrac{k}{2})$. Since the index $k$ in this case is not necessarily restricted to integers, this is actually a continuum of fields.\footnote{Note that the above continuum of positive mass-squared is consistent with the flat limit of the discrete spectrum \eqref{discretespectrum}. Indeed, consider $M^2(\ell)=(\ell-\lambda)(\ell-\lambda+1)/R^2$ for large integer values $\ell\in\dsN_0$ growing to infinity slightly faster than $\lambda$ in such a way that the quotient remains finite. More precisely,
    \begin{align}
        \frac{(\ell-\lambda)(\ell-\lambda+1)}{R^2}\;\xrightarrow[\frac{|\ell-\lambda|}{R}\,\sim\,\scrim(k)>0]{\ell\to\infty,\;\lambda\to\infty,\;R\to \infty}\; M^2(k)\,.
    \end{align}
    } For each given $k\geqslant 0$, the space $\mathcal{W}^{(k)}$ describes a massive unitary irreducible representation (UIR) of $\mathfrak{iso}(1,1)$ with positive squared mass $M^2(k)\geqslant M^2$. In this sense, higher-spin JT gravity obeys the \textit{admissibility condition} of \cite{Konshtein:1988yg}.
    Usually, the UIRs of $\mathfrak{iso}(1,1)$ are considered in the \textit{plane-wave basis} of momentum eigenstates. The explicit map between the twisted-adjoint basis $W^k_{m}$ and such a plane-wave basis can be inferred from its Euclidean analogue discussed in Chapter~9.3 of \cite{Tung}.
    \item An alternative would be a truncation of the matter field by only expanding into linear solutions of \eqref{eq:algebraic_equation_ihs_tadj}; then the result are two decoupled scalar fields ($k=0,1$), both of the same mass $M^2(k)=\scrim^2$.
\end{itemize}

Before concluding, we have to justify the step of switching from dual space to the algebra itself, i.e. from the second equation in \eqref{eqs:covariant_constancy} to \eqref{eq:eom_C}. In the presentation of the AdS case in Subsection~\ref{subsec:higher-spin_BF}, we were able to directly map the fields $C^*$ and $C$ to one another, which is an argument that relies on the semi-simplicity of $\hs[\lambda]$. For the flat case, we already know that the adjoint and coadjoint representations of $\iso(1,1)$ are not equivalent to each other (see also the remarks in Appendix~\ref{app:bilinear_pairing}); however, when the dual space is defined w.r.t. the completion $\widehat{\ihs}[\scrim]$, one can actually intertwine the \emph{twisted}-adjoint and \emph{twisted}-coadjoint representations, which means there exists once again an invertible equivariant mapping that connects $C^*$ with $C$. More comments on that can be found at the end of Appendix~\ref{app:bilinear_pairing}. 
\subsection{Deformed Theory with Backreaction}\label{subsec:backreaction}
As briefly sketched in this subsection, it turns out to be possible to suitably deform the extended higher-spin algebra in order to include a backreaction of the matter fields on the gauge sector. The introduction of non-trivial interactions via a deformation of the associative higher-spin algebra is a general procedure which allows the explicit construction of interaction vertices of any order \cite{Sharapov:2018kjz,Sharapov:2019vyd}. In particular, this procedure was applied in \cite{Bekaert:2025azj} to the higher-spin extension \cite{Alkalaev:2020kut} of JT gravity in the case of non-vanishing cosmological constant. 

An interacting model was proposed in \cite{Bekaert:2025azj} at the level of formal equations of motion; it relies on a deformation of the extended higher-spin algebra $\hs[\lambda]\rtimes\dsZ_2$. This latter deformation admits a natural counterpart in the case of vanishing cosmological constant:
\begin{subequations}
\begin{align}
    & \left[J\,,P_{\pm}\right]=\pm P_{\pm}\,, \qquad \left[P_+\,,P_-\right]=\nu \,\bldtau\,J\,,\\
    & P_+P_-+P_-P_+ -\frac{\nu}{2}\,\bldtau\,=\frac{M^2}2\,\bldidty\,,\\
    & \bldtau\,P_{\pm}=-P_{\pm}\,\bldtau\,, \qquad \bldtau\,J=J\,\bldtau\,, \qquad \bldtau^2=\bldidty\,,
\end{align}
\end{subequations}
where $\nu$ is the deformation parameter. The first line defines the non-trivial deformation of the commutators of $\mathfrak{iso}(1,1)$; the second line fixes the value of the quadratic Casimir element; and the third line specifies the involutive element $\bldtau$ as implementing the twist $\tau$ (as in Section~\ref{subsec:higher-spin_BF}). At $\nu=0$, one recovers the extended higher-spin algebra $\ihs[M]\rtimes\dsZ_2$ of the undeformed theory. 

As explained in \cite{Bekaert:2025azj}, following the general methods from \cite{Sharapov:2018kjz,Sharapov:2019vyd}, the $\nu$-deformation of the extended higher-spin algebra $\hs[\lambda]\rtimes\dsZ_2$ can be converted into formally consistent non-linear equations of motion, which are integrable and whose Lax pair is the flatness and covariant-constancy condition for the deformed algebra. Strictly speaking, the general arguments of \cite{Sharapov:2018kjz,Sharapov:2019vyd} apply to the covariant-constancy condition for the adjoint representation of the deformed extended algebra, while it is the coadjoint representation which is relevant for the zero-form sector when $\Lambda=0$. Nevertheless, we expect that this should not be an obstacle due to the following facts. Firstly, since the deformation of the extended higher-spin algebra $\hs[\lambda]\rtimes\dsZ_2$ admits a non-degenerate invariant metric (which can be constructed from the trace defined in \cite{Bekaert:2025azj}), the adjoint and coadjoint representations are equivalent (modulo, possibly, subtle issues of completion). Therefore, for $\Lambda\neq 0$ the covariant constancy condition can equivalently be written in terms of the adjoint or coadjoint representations. Secondly, the coadjoint representation admits a smooth flat limit $\Lambda\to 0$. Consequently, the vertices associated to flatness and covariant constancy should admit a smooth flat limit when $\Lambda\to 0$. Therefore, the one-parameter family of interacting models from \cite{Bekaert:2025azj} appears to admit a natural analogue when the cosmological constant vanishes. However, we leave this investigation for future work.
\section{Higher-Spin Extension of Cangemi-Jackiw Gravity}\label{sec:higher-spin_CGHS} 
In this section, we explain how the Weyl algebra can be interpreted as an infinite-dimensional higher-spin extension of the Maxwell algebra in two dimensions. Accordingly, we explain how the corresponding BF-type theory can be interpreted as an alternative higher-spin gravity theory in two dimensions, one that extends the $\widehat{\text{CGHS}}$ model.
\subsection{Maxwell and Weyl Algebras}\label{MaxwellWeyl}
As an alternative starting point at spin two, one may consider the non-trivial central extension of the Poincaré algebra $\iso(1,1)$ by the Abelian algebra $\dsR$ generated by a single generator $Z$, say, with the non-vanishing Lie brackets
\begin{align}\label{eq:Maxalg}
    \left[P_+\,,P_-\right]=Z\,, && \left[J\,,P_\pm\right]=\pm P_\pm\,.
\end{align}
Heuristically, this corresponds to considering the trivial central extension $\so(2,1)\oplus\dsR$ of the AdS$_2$ isometry algebra prior to the \.{I}nönü-Wigner contraction and redefining the generators via $P_\pm\to R\,P_\pm$ and $J\to J+R^2Z$ \emph{before taking} the flat limit $R\to\infty$. The algebra described by \eqref{eq:Maxalg} was used in seminal papers such as \cite{Cangemi:1992bj,Nappi:1993ie}. Here, it will be denoted $\mathfrak{ma}(1,1)$ and called the Maxwell algebra (see, e.g., \cite{Schrader:1972zd,Bonanos:2008ez,Gomis:2017cmt}). Physically, this corresponds to a constant magnetic field throughout the Minkowski plane, which effectively renders translations non-commutative.

The Maxwell algebra is the semidirect sum of the Abelian Lorentz subalgebra $\so(1,1)$ acting on the  Heisenberg algebra $\mathfrak{h}_2$ spanned by $P_+$, $P_-$ and $Z$:
\begin{align}
    \mathfrak{ma}(1,1)=\so(1,1)\inplus\mathfrak{h}_2\,.    
\end{align}
In fact, one may easily realise the Maxwell algebra as the Lie algebra spanned by the following four differential operators on a line:
\begin{align}\label{realisationmaxwell}
    P_+=\partial_x\,, && P_-=x\,, && J=-x\frac{\partial}{\partial x}+c-\tfrac{1}{2}\,, && Z=1\,,
\end{align}
with an arbitrary constant $c$. From here one can see two kinds of natural extensions: First, a finite-dimensional extension of the Maxwell algebra is the Schrödinger algebra $\mathfrak{sch}(2)=\mathfrak{sp}(2)\inplus\mathfrak{h}_2$, spanned by all polynomials of degree two in $x$ and $\partial_x$; concretely, one adds two quadratic elements $x^2$ and $\partial^2_x$, and then notes that, together with $x\partial_x$, they form the symplectic algebra $\mathfrak{sp}(2)$ acting on the Heisenberg algebra $\mathfrak{h}_2$. Second, an infinite-dimensional extension of the Maxwell algebra in two dimensions is the Weyl algebra $\mathfrak{A}_2$ of polynomial differential operators in one dimension, spanned by polynomials in $x$ and $\partial_x$. In the present context of two-dimensional gravity, this Weyl algebra $\mathfrak{A}_2$ can be thought of as a higher-spin extension of the Maxwell algebra $\mathfrak{ma}(1,1)$.
\subsection{Cangemi-Jackiw Gravity and its Higher-Spin Generalisations}
In order to investigate the corresponding BF theory, one may start by wondering if there is an invariant non-degenerate metric on the Maxwell algebra. This is a fair question because the Killing form is degenerate since $\mathfrak{ma}(1,1)$ is \emph{not} semisimple (instead, it is solvable). Obviously, the central element $Z$ is a linear Casimir element of the Maxwell algebra. There are two quadratic Casimir elements for the Maxwell algebra, obviously $Z^2$, but also
\begin{align}\label{C2Max}
    \mathcal{C}=\frac{1}{2}\left\{P_+\,,P_-\right\}+JZ\,,
\end{align}
which, in the realisation \eqref{realisationmaxwell}, is proportional to the identity operator, $\mathcal{C}=c\idty$. The Casimir element defines a non-degenerate invariant metric on $\mathfrak{ma}(1,1)$ with non-zero entries
\begin{align}
    \langle P_+\,,P_-\rangle=\langle P_-\,,P_+\rangle=\langle J\,,Z\rangle=\langle Z\,,J\rangle=1\,    
\end{align}
so one can introduce a BF model of the form \eqref{eq:BF_action_bil}. This BF action  has been shown to reproduce the $\widehat{\text{CGHS}}$ model in \cite{Cangemi:1992bj}. 

Note that, while the Maxwell algebra is a quadratic Lie algebra, the Weyl algebra does not admit a non-trivial trace (since any element can be written as a sum of commutators). Again, this does not prevent one from writing a BF action for the Weyl algebra $\mathfrak{A}_2$ in the form \eqref{eq:BF_action_pairing}. This defines a higher-spin extension of CJ gravity in the sense that, if one considers its restriction to the Maxwell subalgebra $\mathfrak{ma}(1,1)\subset\mathfrak{A}_2$, then one recovers CJ gravity \cite{Cangemi:1992bj}. Indeed, since Maxwell algebra is quadratic, the form \eqref{eq:BF_action_bil} and \eqref{eq:BF_action_pairing} of the corresponding BF action are equivalent. 
 
Actually, one can say more because the Weyl algebra is known to admit a (unique) \emph{supertrace} \cite{Vasiliev:1995dn}.  As a first comment, let us emphasise that the abstract algebra constructions provided in Subsection \ref{subsec:higher-spin_BF} for the example of the higher-spin algebra $\hs[\lambda]\supset\so(2,1)$ can also be applied to the Weyl algebra $\mathfrak{A}_2\supset\mathfrak{ma}(1,1)$. In fact, the twist operation defined by $\tau(J)=J$, $\tau(Z)=Z$ and $\tau(P_\pm)=-P_\pm$ is an involutive automorphism of the Maxwell algebra. This automorphism straightforwardly extends to the Weyl algebra since it can be implemented as the reflection $\tau:x\mapsto -x$, as can be checked on the realisation \eqref{realisationmaxwell}. Being involutive, the twist defines a $\dsZ_2$-grading of the Weyl algebra, which decomposes as the direct sum, $\mathfrak{A}_2=\mathfrak{A}_2^+\oplus\mathfrak{A}_2^-$, of the eigenspaces $\mathfrak{A}_2^\pm$ with $\tau$-eigenvalue $\pm 1$. In this sense, the Weyl algebra is a superalgebra.

A standard realisation of the Weyl algebra $\mathfrak{A}_2$ is as the space $\dsR[x_1,x_2]$ of polynomial symbols, i.e. functions on the plane with coordinates $x_\alpha=(x_1,x_2)$ endowed with the Moyal-Groenewold product
\begin{align}
    \big(f*g\big)(x)=\left[\exp
    \left(\tfrac12\,\varepsilon_{\alpha\beta}\frac{\partial}{\partial y_\alpha}\frac{\partial}{\partial z_\beta}\right)f(y)g(z)\right]_{z=y=x}\,.
\end{align}
In this realisation, the twist is the operation $\tau:x_\alpha\to-x_\alpha$ which is indeed an automorphism of the Weyl algebra, $\tau(f*g)=\tau(f)*\tau(g)$. One can define the supertrace $\operatorname{str}(f):=f(0)$ as the evaluation at the origin of the plane. This defines a linear form $\operatorname{str}:\mathfrak{A}_2\to\dsR$ that (i) annihilates $\mathfrak{A}_2^-$ and (ii) obeys the graded-cyclicity property 
\begin{align}
    \operatorname{str}(f_\pm\star g_\pm)=\pm\,\operatorname{str}(g_\pm\star f_\pm)\,,\qquad\operatorname{str}(f_\pm\star g_\mp)=0\,,
\end{align}
for any $f_\pm,g_\pm\in\mathfrak{A}_2^\pm$. The properties (i) and (ii) are equivalent to the two following properties: the supertrace is $\tau$-invariant, $\operatorname{str}\big(\tau(f)\big)=\operatorname{str}\big(f\big)$, and it obeys the twisted-cyclicity property
\begin{align}
    \operatorname{str}\big(f*g\big)=\operatorname{str}\big(g *\tau(f)\big)\,.
\end{align}
In turn, one can introduce a non-degenerate graded-symmetric bilinear map on the Weyl algebra via the supertrace
\begin{align}\label{nondegbilformA2}
    \langle f,g\rangle:=\operatorname{str}\big(f*\tau(g)\big)=\operatorname{str}\big(g*f\big)\,.
\end{align}
The Moyal bracket is the commutator w.r.t. the Moyal-Groenewold product, 
\begin{equation}
[f,g]_*(z):=(f*g)(z)-(g*f)(z)=2\left[\sinh
    \left(\tfrac12\,\varepsilon_{\alpha\beta}\frac{\partial}{\partial y_\alpha}\frac{\partial}{\partial z_\beta}\right)f(y)g(z)\right]_{z=y=x}\,.    
\end{equation}
It reproduces the Lie bracket of $\mathfrak{ma}(1,1)$ under the identification
\begin{align}\label{eqs:Maxwell_plane}
     P_+=x_1\,, && P_-=x_2\,, && J=-x_1x_2+c\,, && Z=1\,.
\end{align}
Unfortunately, the graded-symmetry of the bilinear form \eqref{nondegbilformA2} implies that it does not restrict to the metric on the Maxwell algebra; in fact, the non-zero entries of the bilinear form read
\begin{align}
    \langle Z\,,Z\rangle=1\,, && \left\langle P_+\,,P_-\right\rangle=-\left\langle P_-\,,P_+\right\rangle=-\frac{1}{2}\,, && \langle Z\,,J\rangle=\langle J\,,Z\rangle=c\,, && \langle J\,,J\rangle=c^2-\frac{1}{4}\,.
\end{align}

However, the non-degenerate bilinear form \eqref{nondegbilformA2} defines an injective intertwiner $\Phi:\mathfrak{A}_2\to\mathfrak{A}_2^*$ relating the twisted-adjoint and the coadjoint representations of the Weyl algebra \cite{Skvortsov:2022syz} since
\begin{align}
    \left\langle f\,,\ad_h(g)\right\rangle=-\left\langle\tad{_h}(f)\,,\,g\right\rangle\,.
\end{align}
Accordingly, it defines a BF action \eqref{eq:BF_action_bil} where $A$ and $B$ both take values in the Weyl algebra $\mathfrak{A}_2$ but live in the adjoint and twisted-adjoint representation, respectively. By construction, this BF theory should be equivalent to the one based on the pairing, i.e. \eqref{eq:BF_action_pairing}, mentioned above.  

Endowed with all these preliminary facts about the Weyl algebra, one may now consider a further extension of CJ gravity by considering the smash product $\mathfrak{A}_2\rtimes\dsZ_2$, where $\dsZ_2=\{\idty,\tau\}$. This extended algebra is particularly appealing because, as an associative algebra, it admits a unique deformation,  
\begin{equation}\label{defosc}
    x_\alpha\star x_\beta-x_\beta\star x_\alpha=\epsilon_{\alpha\beta}\,(\bldidty+\nu\,\bldtau)\,,\quad x_\alpha\star\bldtau=- \bldtau\star x_\alpha\,,\quad \bldtau\star\bldtau=\bldidty\,,
\end{equation}
usually called the \textit{deformed oscillator algebra} \cite{Vasiliev:1989re} (see e.g. Section~6.1 in \cite{Sharapov:2018kjz} for a concise review). Applying the general arguments of \cite{Sharapov:2018kjz,Sharapov:2019vyd} to the present case, one finds that the deformed oscillator algebra defines a formal higher-spin gravity in two dimensions with non-trivial interactions (in particular, couplings between the gauge and matter sectors) extending CJ gravity. The extended algebra $\mathfrak{A}_2\rtimes\dsZ_2$ is also appealing because it is endowed with a trace, thereby allowing the construction of a BF action principle in its traditional form \eqref{eq:BF_action_tr}.

The elements of $\mathfrak{A}_2\rtimes\dsZ_2$ are pairs $(f,g)$ with $f,g\in\mathfrak{A}_2$ that can be written $f\bldidty+g\bldtau$ and obey the multiplication rule (corresponding to \eqref{defosc} at $\nu=0$)
\begin{align}
    \left(f_1\bldidty+g_1\bldtau\right)\star\left(f_2\bldidty+g_2\bldtau\right)=\big(f_1*f_2+g_1*\tau(g_2)\big)\bldidty+\big(f_1*g_2+g_1*\tau(f_2)\big)\bldtau\,.
\end{align}   
In a suitable completion of the Weyl algebra, it admits a realisation as elements of the form $f+g*\delta$ and multiplication rule
\begin{align}
    \left(f_1+g_1*\delta\right)*\left(f_2+g_2*\delta\right)=\big(f_1* f_2+g_1*\tau(g_2)\big)+\big(f_1*g_2+g_1*\tau(f_2)\big)*\delta\,,
\end{align}
where the Dirac distribution $\delta(x):=\delta(x_1)\delta(x_2)$ obeys $\delta*f=\tau(f)*\delta$ (as can be proved by checking it on the generating function of all monomials in $x$, i.e. on $f(x_\alpha)=e^{y^\alpha x_\alpha}$).

In any case, one can define a trace $\tr:\mathfrak{A}_2\rtimes\dsZ_2\to\dsR$ on the extended algebra as
\begin{equation}
    \tr(f\,\bldidty+g\,\bldtau)=\operatorname{str}(g)
\end{equation}
and check that it indeed obeys the cyclicity property
\begin{subequations}
\begin{align}
    \tr\Big(\left(f_1\bldidty+g_1\bldtau\right)\star\left(f_2\bldidty+g_2\bldtau\right)\Big)&=
        \operatorname{str}\Big(f_1*g_2+g_1*\tau(f_2)\Big)=\operatorname{str}\Big(f_2*g_1+g_2*\tau(f_1)\Big)\\
    &=\tr\Big(\left(f_2\bldidty+g_2\bldtau\right)\star\left(f_1\bldidty+g_1\bldtau\right)\Big)\,.
\end{align}
\end{subequations}
Accordingly, we propose as higher-spin extension of CJ gravity the BF action for the corresponding trace, where $A$ and $B$ both take values in the extended higher-spin algebra $\mathfrak{A}_2\rtimes\dsZ_2$. 

Let us stress that this long digression using symbol calculus is not necessary and one may as well work only in the realisation of the Weyl algebra as polynomial differential operators in one dimension. For instance, $\bldtau$ identifies with $(-1)^J=\exp(i\pi J)$ and one can explicitly see the relation between the operators $\hat A(x,\partial_x)$ on the line and symbols $A(x_1,x_2)$ on the plane in terms of the Weyl map 
\begin{align}
    \hat{A}\mapsto A
    =e^{\frac12\partial_x\partial_y}\left(e^{-xy}\cdot\hat{A}\cdot e^{xy}\,\right)_{y=x_1,\,x=x_2}\,.
\end{align}
Accordingly, one recovers \eqref{eqs:Maxwell_plane} from \eqref{realisationmaxwell}.
\subsection{Higher-Spin Gravity: Jackiw-Teitelboim vs. Cangemi-Jackiw}
One may interpret the higher-spin algebra $\ihs[M]$, i.e. the quotient \eqref{ihsM}, as the algebra of functions on the fuzzy plane (by analogy with the comment in Footnote~\ref{noncomgeom}). In fact, a suitable completion of $\ihs\left[\scrim\right]$ is isomorphic to (a completion of) the Weyl algebra $\mathfrak{A}_2$, as can be seen from the realisation $P_\pm=\scrim\exp(\pm x)$ and $J=\partial_x$ of the generators as differential operators. This isomorphism is also consistent with the fact the Maxwell algebra $\mathfrak{ma}(1,1)$ can be obtained from an \.{I}nönü-Wigner contraction of the trivial central extension $\mathfrak{so}(2,1)\oplus\mathfrak{u}(1)$ with a redefinition of the boost generator before taking the flat limit (cf. Subsection~\ref{MaxwellWeyl}). Although some completions of the algebras $\ihs[M]$ and $\mathfrak{A}_2$ are isomorphic, this does not imply that they necessarily define the same higher-spin gravity theories.

More generally, the isomorphism of the underlying algebras $\g$ of two BF models does \textit{not} necessarily imply the equivalence of the corresponding gravity theories, because a crucial physical input for a gravitational interpretation is the choice of a spin-two subalgebra, e.g. $\iso(1,1)\subset\g$ or $\mathfrak{ma}(1,1)\subset\g$.  Consequently, one has to specify what the translation generators are. This is a necessary requirement for a gravitational interpretation of a BF model, because the corresponding zweibein must be non-degenerate, $\det(e_\mu^a)\neq 0$. For instance, this implies that flat connections cannot be gauged away, although they are formally pure gauge. The spectrum of a BF model with a gravitational interpretation is obtained by picking a spin-two background (a flat connection $A_0$ and a covariantly constant dilaton $B_0$, both taking values in the spin-two subalgebra) and analysing the solutions of the linearised equations of motion. Let us stress that the higher-spin gravity theories with vanishing cosmological constant considered in Sections \ref{sec:flat_case} and \ref{sec:higher-spin_CGHS} are a priori two inequivalent theories, since their spin-two background solutions will differ and, accordingly, their spectrum will be organised very differently.

We leave the computation of the spectrum of the higher-spin extension of CJ gravity for further study.  Following the standard recipe in higher-spin theory, this computation would require to decompose the adjoint and twisted-adjoint representation of $\mathfrak{ma}(1,1)$ on $\mathfrak{A}_2$ as a sum of irreducible $\mathfrak{ma}(1,1)$-modules, which is a non-trivial task. In particular, one should verify that the analogue of the admissibility condition \cite{Konshtein:1988yg} on higher-spin gravity theories is fulfilled by our extension of CJ gravity: the twisted-adjoint representation carried by $\mathfrak{A}_2$ should decompose into a sum of UIRs of $\mathfrak{ma}(1,1)$. Fortunately, the representation theory of the latter Maxwell algebra is already known \cite{deMello:2002zkl}.
\section{Discussion}\label{sec:discussion}
In this work, we demonstrated how the BF action can be written in a way suited to describe both JT gravity and its various higher-spin generalisations at zero cosmological constant. Issues related to the non-existence of an invariant bilinear form on Lie algebras containing an Abelian ideal are circumvented by re-writing the action functional in terms of the pairing between the respective algebra and its dual. In particular, the dilaton zero-form takes values in the dual of the Lie algebra and the connection one-form acts on it via the coadjoint action.

We defined some flat-space higher-spin algebras from which we constructed higher-spin gravity theories in two dimensions with vanishing cosmological constant. In particular, we mentioned a suitable \.{I}nönü-Wigner contraction of $\sl(N,\dsR)$ which provides examples of flat-space finite-dimensional higher-spin algebras. We furthermore defined the flat-space infinite-dimensional higher-spin algebra $\ihs[\scrim]$ and its extension $\ihs[\scrim]\rtimes\dsZ_2$. The latter is needed in order to introduce massive scalar degrees of freedom to the otherwise topological gauge theory and it is constructed by means of an involutive automorphism on the associative algebra, giving rise to what is called the twisted-adjoint action. We showed that the extended BF action describes an infinite, continuous tower of scalar fields of varying mass, transforming in the respective twisted-adjoint representation. It is an important difference to the analogous case in AdS$_2$ that this tower is not discrete. Another difference lies in the fact that, while $\hs[\lambda]$ reduces to the finite-spin algebras $\sl(N,\dsR)$ for integer values $\lambda=N$ of the parameter, this is not true for the Lie algebra $\ihs[\scrim]$, which does not contain any non-trivial ideal for special values $\scrim^2$ of the quadratic Casimir element (with the obvious exception of $\scrim=0$). So the relation between the infinite-dimensional and the finite-dimensional higher-spin algebras seems more subtle for $\Lambda=0$.

Furthermore, we have shown that it is possible to suitably deform the extended higher-spin algebra $\ihs[\scrim]\rtimes\dsZ_2$ in order to include a backreaction of the matter fields on the gauge sector. This proposal is based on \cite{Bekaert:2025azj} where two models, A and B, of two-dimensional interacting higher-spin gravity theories with non-vanishing cosmological constant were proposed. The model A holds at the level of formal equations of motion and relies on a deformation of the extended higher-spin algebra. As argued in Section~\ref{subsec:backreaction}, Model A admits a natural analogue for $\Lambda=0$. It would be interesting to understand if the same is true for Model B (since the latter admits an action principle in the form of a Poisson sigma model).

Finally, a higher-spin extension of CJ gravity was proposed in Section~\ref{sec:higher-spin_CGHS} by replacing the Maxwell algebra $\mathfrak{ma}(1,1)$ with the Weyl algebra $\mathfrak{A}_2$. Similar considerations as in the case of $\ihs[\scrim]$ were applied to the case of $\mathfrak{A}_2$. In particular, we pointed out the nice algebraic properties of the extension $\mathfrak{A}_2\rtimes\dsZ_2$ ensuring that this higher-spin theory allows for a standard BF action and for a non-trivial deformation. We left the analysis of its spectrum for future study.

There are several possibilities to generalise or extend the present work: 
\begin{itemize}
    \item It would be interesting to consider a supersymmetric extension \cite{Teitelboim:1983uy,Montano:1990ru,Cardenas:2018krd} of the Poincaré algebra and investigate the corresponding BF-type theory.
    \item There exists a coloured version of JT gravity \cite{Alkalaev:2022szj} corresponding to the extension $\mathfrak{su}(N,N)$ of the isometry algebra $\so(2,1)\simeq\mathfrak{su}(1,1)$. A natural question is whether its flat limit (obtained by Abelianising the generalised translations) admits a nice gravitational interpretation.
    \item It is known \cite{Hoppe:1982,Bergshoeff:1989ns} that the following limits of higher-spin algebras coincide with the Poisson algebra $w_\infty$ of functions on the hyperboloid (aka. the area-preserving diffeomorphism algebra):
    \begin{align}
        \lim\limits_{N\to\infty}\mathfrak{sl}(N,\dsR)=\lim\limits_{\lambda\to\infty}\hs[\lambda]=w_\infty\,.
    \end{align}
    This raises the question whether the flat limit commutes with the above limits and simply gives the Poisson algebra $\mathcal{P}$ of functions on the plane. In particular, it would be nice to check explicitly whether the following commutative square holds
    \begin{equation*}
    \begin{tikzcd}[column sep=12em, row sep=5em]
        \hs[\lambda] \arrow[r, "\text{\.{I}nönü-Wigner contraction}", "R\to\infty"'] \arrow[d,"\lambda\to\infty", "\text{Classical limit}"'] & \ihs[M] \arrow[d,  "M\to\infty","\text{Classical limit}"'] \\
        w_\infty \arrow[r, "\text{\.{I}nönü-Wigner contraction}", "R\to\infty"'] & \mathcal{P}
    \end{tikzcd}     
    \end{equation*}
    and investigate these limits for the corresponding higher-spin gravity theories.
    \item As another possible direction of further investigations, it would be pleasant to make the interacting Model A alluded above more concrete by trying to extract explicit interaction vertices. The flat version proposed here might be of interest since one  expects the vertices to take a simpler form when the cosmological constant vanishes. 
    \item The CGHS model is a useful playground for exploring black holes in two dimensions, thus it would be nice to look for solutions of black-hole type for its higher-spin extension. 
    \item It would also be interesting to investigate the ``classical'' limit of the higher-spin extension of CJ gravity when the Weyl algebra $\mathfrak{A}_2$ becomes the Poisson algebra $\mathcal P$ mentioned above. 
    \item We have not attempted to discuss our flat-space higher-spin proposals from the viewpoint of holography, so far. This would necessitate to choose suitable boundary conditions and include an appropriate boundary term to the BF action, in order to study the asymptotic symmetries of the model as in \cite{Gonzalez:2018enk}. In CJ gravity, specific boundary conditions have been introduced \cite{Afshar:2019axx}, leading however to an asymptotic symmetry that is of warped-Virasoro type rather than $\mathfrak{bms}_2$ (for remarks on the latter, see \cite{Afshar:2021qvi}). Interestingly, a matrix model dual to CJ gravity has been proposed \cite{Kar:2022sdc}.
    \item Based on BF actions, a higher-spin generalisation of the Saad-Shenker-Stanford duality was proposed in \cite{Kruthoff:2022voq}, while supersymmetric localisation of higher-spin JT gravity theories was investigated in \cite{Griguolo:2023aem}. The possibility of flat-space counterparts to these works remains tantalising.
\end{itemize}
We leave these various questions open for future research.
\section*{Acknowledgements}
The research of MP is supported by the European Union\footnote{Views and opinions expressed are however those of the author only and do not necessarily reflect those of the European Union or the European Research Council. Neither the European Union nor the granting authority can be held responsible for them.} (ERC grant ``HoloBoot'', project number 101125112), by the MUR-PRIN grant No.\,PRIN2022BP52A (European Union -- Next Generation EU) and by the INFN initiative STEFI.
%
%
%
%
%
%
%
%
%
\appendix
\addtocontents{toc}{\protect\setcounter{tocdepth}{1}}
\section{On Bilinear Forms and the Pairing}\label{app:bilinear_pairing}
In this appendix we would like to illustrate the possibility to define a bilinear form on a Lie algebra $\g$ in terms of a mapping $\Phi:\g\rightarrow\g^*$ from the algebra to its dual space $\g^*$. We will consider the semisimple case $\g=\so(2,1)$ as well as the non-semisimple case $\g=\iso(1,1)$; additionally, we remark on the interesting case $\g=\widehat{\ihs}[\scrim]$.
\subsection{The Semisimple Case}
For the algebra $\so(2,1)$ with Lie brackets \eqref{eq:algebra_generalL} the relations are straightforward. A basis of the dual algebra is defined in terms of a pairing,
\begin{align}\label{dualbasis}
    \left(J^*,J\right):=1\,, && \left(P^{*a}\,,P_b\right):=\delta^a_b\,, && \text{zero otherwise.}
\end{align}
Then the invertible linear mapping $\Phi:\,\so(2,1)\rightarrow\so(2,1)^*$ is defined as
\begin{align}
    \Phi(J):=J^*\,, && \Phi(P_a):=-\Lambda\eta_{ab}P^{*b}\,,
\end{align}
such that the bilinear form on $\so(2,1)$ can be defined as $\langle X\,,Y\rangle:=(\Phi(X)\,,Y)$ and agrees with the adjoint-invariant one presented in Subsection~\ref{subsec:classical_grav}. Note that the cosmological constant is present in the mapping in order to account for the length dimension of transvection generators.
\subsection{The Non-Semisimple Case}
\minisec{The Case of Poincaré}
For the Poincaré algebra $\iso(1,1)$ with Lie brackets \eqref{eq:Lie_bracket_Poincare} the definition of a basis of the dual algebra is exactly the same as above, namely \eqref{dualbasis}. Let us define a linear mapping $\Phi:\,\iso(1,1)\rightarrow\iso(1,1)^*$ similarly as before,
\begin{align}\label{linearmapPhi}
    \Phi(J):=J^*\,, && \Phi(P_a):=\scrim^2\eta_{ab}P^{*b}\,,
\end{align}
where the factor $\scrim^2$ has been included in order to compensate for the length dimension contained in translation generators, which is the same logic as in the case of $\so(2,1)$. In the main text, we did not include that factor since, due to the commutativity of translations, their generators can be freely rescaled. Again, the bilinear form on $\iso(1,1)$ can be defined as $\langle X\,,Y\rangle:=(\Phi(X)\,,Y)$, which results in
\begin{align}\label{bilinearform}
    \left\langle J\,,J\right\rangle=1\,, && \left\langle P_a\,,P_b\right\rangle=\scrim^2\eta_{ab}\,, && \text{zero otherwise.}
\end{align}
Up to now, this is nothing but a redefinition $\Lambda\to -M^2$ of the previous case.
However, the crucial point is that the non-degenerate bilinear form \eqref{bilinearform} is \textit{not} invariant under the adjoint action of $\iso(1,1)$ since
\begin{align}\label{noninvariance}
  0=\left\langle\, [P_a,P_b]\,,J\,\right\rangle\neq\left\langle P_a\,,[P_b,J]\right\rangle=M^2\varepsilon_{ab}\,.
\end{align}
In a sense, the main insight of Subsection~\ref{subsec:classical_grav} is that it is not necessary to be in possession of a fully ad-invariant bilinear form in order to write down a sensible BF theory (the non-degeneracy is necessary in order to have as many equations of motion as there are fields). For the case at hand of JT gravity with vanishing cosmological constant, it is indeed possible to define a gauge invariant BF action in terms of a non-degenerate bilinear form that is not adjoint-invariant, such as \eqref{bilinearform}. This is possible because the gauge symmetries (and equations of motion) of the zero-form are not the naive adjoint action on $B$, they are modified and only take a natural form in terms of the coadjoint action on $B^*=\Phi(B)$.

Let us stress that there is no contradiction with the no-go theorem on the invariant bilinear form because the invertible linear mapping $\Phi:\iso(1,1)\rightarrow\iso(1,1)^*$ defined by \eqref{linearmapPhi} does \textit{not} intertwine the adjoint and coadjoint representations, as can be checked explicitly from the counterpart of \eqref{noninvariance}:
\begin{align}\label{differentfrom}
  0=\big(\,\Phi\cdot[P_a,P_b]\,,J\,\big)\neq\big(\Phi\cdot P_a\,,[P_b,J]\big)=M^2\varepsilon_{ab}\,.
\end{align}
If we had an equality in the middle of \eqref{differentfrom}, then $\Phi$ would be an intertwiner relating the adjoint and coadjoint representations of $\iso(1,1)$. However, such an invertible intertwiner $\Phi:\iso(1,1)\rightarrow\iso(1,1)^*$ \textit{cannot} exist, because it would define a non-degenerate ad-invariant bilinear form on  $\iso(1,1)$ (which  cannot exist, as recalled in Subsection~\ref{subsec:classical_grav}). In other words, the adjoint and coadjoint representations are \textit{inequivalent} representations for $\iso(1,1)$, while they are equivalent representations for $\so(2,1)$. This is another way to understand why the more general form of the BF action is necessary in order to write a variational principle for JT gravity with zero cosmological constant in the gauge formulation.
\minisec{The Case of $\widehat{\ihs}[\scrim]$}
Though the argumentation just described applies in the same way to the higher-spin extension $\ihs[\scrim]$ of $\iso(1,1)$, the situation changes drastically once a suitable completion of that former vector space is considered. Recall that we define the said completion explicitly as the span of the generators $W^k_{\pm m}$ in Subsection~\ref{subsec:ihs_tw_basis}: the latter are defined in terms of a nested application of the twisted-adjoint action of translations in \eqref{eq:def_tadj_generators}. In that case, we define the respective dual space $\widehat{\ihs}[\scrim]^*$ as the span of generators $W^{*k}_{\pm m}$ with defining property
\begin{align}
    \left(W^{*k}_m\,,W^{k'}_{m'}\right):=\delta^{kk'}\delta_{m+m',0}\,.
\end{align}
It is then easy to see that this definition is consistent with the existence of an invertible mapping
\begin{align}
    \Phi:\ \widehat{\ihs}[\scrim]\rightarrow\widehat{\ihs}[\scrim]^*\,,\ \ W^k_m\mapsto\Phi\cdot W^k_m:=W^{*k}_m
\end{align}
that intertwines the twisted-adjoint representation and the twisted-coadjoint representation.
This follows directly from $\{P_\pm\,,W^k_{\pm m}\}=W^k_{\pm(m+1)}$ and $\{P_\mp\,,W^k_{\pm m}\}=M^2(k)W^k_{\pm(m-1)}$, giving
\begin{subequations}
\begin{align}
    M^2(k)\left(\Phi\cdot\tad{_\pm}\big(W^k_{\pm m}\big)\,,W^{k'}_{\mp m'}\right)&=\left(\Phi\cdot W^k_{\pm m}\,, \tad{_\pm}\big(W^{k'}_{\mp m'}\big)\right)\,, && m\in\dsN_0\,,\ m'\in\dsN\\
    \left(\Phi\cdot\tad{_\pm}\big(W^k_{\mp m}\big)\,,W^{k'}_{\pm m'}\right)&=M^2(k)\left(\Phi\cdot W^k_{\mp m}\,, \tad{_\pm}\big(W^{k'}_{\pm m'}\big)\right)\,, && m\in\dsN\,,\ m'\in\dsN_0\,,
\end{align}
\end{subequations}
while the relation for the action of $J$ is straightforward. In conclusion, we deduce from this intertwiner that it is actually possible to define a \emph{twisted}-adjoint-invariant non-degenerate bilinear form on a suitable \emph{completion} of the flat-space higher-spin algebra. 

The same logic would apply to the adjoint basis and the respective vector-space completion spanned by generators $V^s_m$ in Subsection~\ref{subsec:ihs_tw_basis}. Nevertheless, there is no contradiction with the fact that $\iso(1,1)$ does not admit an adjoint-invariant, non-degenerate, bilinear form. In fact, keep in mind that $\{V^s_m\}$ is \emph{not} a highest-weight basis, meaning the index $m$ is not truncated and the adjoint action of $\iso(1,1)$ on these generators never gives zero, this avoids a contradiction of the form \eqref{differentfrom}. 

The  lesson learned is that we can \emph{either} define a twisted-adjoint-invariant bilinear form on a completion of $\ihs[\scrim]$ \emph{or} an adjoint-invariant bilinear form on a different completion of $\ihs[\scrim]$ which are non-degenerate but not both at the same time, i.e. in particular not on the extended higher-spin algebra $\ihs[\scrim]\rtimes\dsZ_2$.
\section{Systematic Approach to the Algebraic Equation}\label{app:sec:systematic_app}
In the main text, there are different occasions on which we encountered `algebraic' equations, all being of the following type:
\begin{align}\label{eq:general_equation}
    2\left(c+\Lambda x^2\right)f(x)\mp\left(c+\Lambda x(x+1)\right)f(x+1)\mp\left(c+\Lambda x(x-1)\right)f(x-1)=\mu f(x)\,.
\end{align}
This expression contains all four cases: the adjoint basis for the upper and the twisted-adjoint basis for the lower sign; AdS for $\Lambda=-1$ as well as $c=(\lambda^2-1)/4$, and flat space for $\Lambda=0$ as well as $c=\scrim^2/4$. Here we present a possible approach to study such an equation.

The idea is to consider the discretised version of the problem \eqref{eq:general_equation}, meaning that we replace $x\mapsto n\in\dsN_0$ and $f(x)\mapsto a_n$. We are then dealing with the recurrence relation
\begin{align}\label{eq:general_equation_discrete}
        2\left(c+\Lambda n^2\right)a_n\mp\left(c+\Lambda n(n+1)\right)a_{n+1}\mp\left(c+\Lambda n(n-1)\right)a_{n-1}=\mu a_n\,.
\end{align}
Defining the generating function
\begin{align}
    F(z)=\sum_{n=0}^\infty a_n z^n
\end{align}
we obtain the differential equation (here abbreviating $\tilde{\mu}\equiv-1+\mu/(2c)$)
\begin{align}\label{eq:general_equation_generating}
    \frac{\Lambda}{c}z^2(z\mp 1)\left((z\mp 1)F''(z)+2F'(z)\right)+\left(z^2\pm 2\tilde{\mu}z+1\right)F(z)=a_0+\left(a_1\pm 2\tilde{\mu}a_0\right)z\,.
\end{align}
In the following, let us consider the case $\Lambda=0$, only.

At vanishing cosmological constant, equation \eqref{eq:general_equation_generating} becomes algebraic and its left-hand side vanishes completely at the special values $z_1=\mp\tilde{\mu}+\sqrt{\tilde{\mu}^2-1}$ and $z_2=\mp\tilde{\mu}-\sqrt{\tilde{\mu}^2-1}$, which we exclude for the moment. Then we can write
\begin{align}
    F(z)=\frac{a_0+\left(a_1\pm 2\tilde{\mu}a_0\right)z}{z^2\pm2\tilde{\mu}z+1}\,,
\end{align}
which under the assumption $z_1\ne z_2$, i.e. $\tilde{\mu}^2\ne 1$, can be decomposed into partial fractions, resulting in
\begin{align}
    F(z)=\frac{\alpha_1}{1-\frac{z}{z_1}}+\frac{\alpha_2}{1-\frac{z}{z_2}}=\alpha_1\sum_{n=0}^\infty z_1^{-n}z^n+\alpha_2\sum_{n=0}^\infty z_2^{-n}z^n\,,
\end{align}
where $\alpha_{1,2}\equiv (z_{1,2}a_0-a_1)/(z_{1,2}-z_{2,1})$. So, in the continuous generalisation, what we found are solutions of exponential type as presented in the main part. In particular, setting $z_1=\e{k}$ and $z_2=\e{-k}$ we find
\begin{align}
    \mu=2c\left(1\mp\frac{z_1+z_2}{2}\right)=\begin{cases}
        -4c\sinh^2\left(\frac{k}{2}\right)\,,   & \text{adjoint}\,;\\
        4c\cosh^2\left(\frac{k}{2}\right)\,,    & \text{twisted-adjoint}\,.
    \end{cases}
\end{align}

In addition, the special cases $\tilde{\mu}=-1$ ($\mu=0$) and $\tilde{\mu}=1$ ($\mu=4c$) need to be considered. The former results in\footnote{Note that both the upper sign (adjoint case) at $\mu=0$ and the lower sign (twisted-adjoint case) at $\mu=4c$ result in the arithmetic sequence, any member of which is given as the arithmetic mean of its neighbours.}
\begin{align}
    F(z)=\big(a_0-(a_0\mp a_1)z\partial_z\big)\frac{1}{1\mp z}=\sum_{n=0}^\infty (\pm 1)^n\big(a_0-(a_0\mp a_1)n\big)z^n
\end{align}
and the latter in
\begin{align}
    F(z)=\big(a_0-(a_0\pm a_1)z\partial_z\big)\frac{1}{1\pm z}=\sum_{n=0}^\infty (\mp 1)^n\big(a_0-(a_0\pm a_1)n\big)z^n\,.
\end{align}
Transferring to the continuous case, we found the linear solutions, $f(x)=ax+b$, as well as linear solutions with a phase factor, $f(x)=\e{\i\pi x}(ax+b)$.

Finally, coming back to the special values $z=z_{1,2}$ that have been excluded above, one can check that, although a different generating function appears after fixing the boundary values $a_0$ and $a_1$, no additional solutions are obtained in that way.

Although the formal transition to the discrete level and back may be somewhat questionable, the rather structured approach in terms of generating functions appears reassuring that the solutions which were guessed in the main text should actually cover all possibilities.
\printbibliography

@article{Kar:2022sdc,
    author = "Kar, Arjun and Lamprou, Lampros and Marteau, Charles and Rosso, Felipe",
    title = "{A Matrix Model for Flat Space Quantum Gravity}",
    eprint = "2208.05974",
    archivePrefix = "arXiv",
    primaryClass = "hep-th",
    doi = "10.1007/JHEP03(2023)249",
    journal = "JHEP",
    volume = "03",
    pages = "249",
    year = "2023"
}

@article{Nappi:1993ie,
    author = "Nappi, Chiara R. and Witten, Edward",
    title = "{A WZW model based on a nonsemisimple group}",
    eprint = "hep-th/9310112",
    archivePrefix = "arXiv",
    reportNumber = "IASSNS-HEP-93-61",
    doi = "10.1103/PhysRevLett.71.3751",
    journal = "Phys. Rev. Lett.",
    volume = "71",
    pages = "3751--3753",
    year = "1993"
}

@article{deMello:2002zkl,
    author = "de Mello, R. O. and Rivelles, Victor O.",
    title = "{The Irreducible unitary representations of the extended Poincare group in (1+1)-dimensions}",
    eprint = "math-ph/0208024",
    archivePrefix = "arXiv",
    doi = "10.1063/1.1644901",
    journal = "J. Math. Phys.",
    volume = "45",
    pages = "1156--1167",
    year = "2004"
}

@article{Konshtein:1988yg,
    author = "Konshtein, S. E. and Vasiliev, Mikhail A.",
    title = "{Massless Representations and Admissibility Condition for Higher Spin Superalgebras}",
    reportNumber = "LEBEDEV-88-94",
    doi = "10.1016/0550-3213(89)90301-5",
    journal = "Nucl. Phys. B",
    volume = "312",
    pages = "402--418",
    year = "1989"
}

@article{Vasiliev:1989re,
    author = "Vasiliev, Mikhail A.",
    title = "{Higher Spin Algebras and Quantization on the Sphere and Hyperboloid}",
    reportNumber = "LEBEDEV-89-214",
    doi = "10.1142/S0217751X91000605",
    journal = "Int. J. Mod. Phys. A",
    volume = "6",
    pages = "1115--1135",
    year = "1991"
}

@article{Vasiliev:1995dn,
    author = "Vasiliev, Mikhail A.",
    editor = "Berezin, V. A. and Rubakov, V. A. and Semikoz, D. V.",
    title = "{Higher spin gauge theories in four-dimensions, three-dimensions, and two-dimensions}",
    eprint = "hep-th/9611024",
    archivePrefix = "arXiv",
    reportNumber = "FIAN-TD-24-96",
    doi = "10.1142/S0218271896000473",
    journal = "Int. J. Mod. Phys. D",
    volume = "5",
    pages = "763--797",
    year = "1996"
}

@article{Skvortsov:2022syz,
    author = "Skvortsov, Evgeny and Van Dongen, Richard",
    title = "{Minimal models of field theories: Chiral higher spin gravity}",
    eprint = "2204.10285",
    archivePrefix = "arXiv",
    primaryClass = "hep-th",
    doi = "10.1103/PhysRevD.106.045006",
    journal = "Phys. Rev. D",
    volume = "106",
    number = "4",
    pages = "045006",
    year = "2022"
}

@article{Sharapov:2018kjz,
    author = "Sharapov, Alexey and Skvortsov, Evgeny",
    title = "{$A_\infty$ algebras from slightly broken higher spin symmetries}",
    eprint = "1809.10027",
    archivePrefix = "arXiv",
    primaryClass = "hep-th",
    doi = "10.1007/JHEP09(2019)024",
    journal = "JHEP",
    volume = "09",
    pages = "024",
    year = "2019"
}

@book{Tung,
author = {Tung, Wu-Ki},
title = {Group Theory in Physics},
publisher = {World Scientific},
year = {1985},
doi = {10.1142/0097},
}

@phdthesis{Hoppe:1982,
  author       = {Hoppe, Jens},
  title        = {Quantum Theory of a Massless Relativistic Surface and a Two-Dimensional Bound State Problem},
  school       = {Massachusetts Institute of Technology},
  year         = {1982},
  address      = {Cambridge, MA},
  url          = {https://dspace.mit.edu/handle/1721.1/15717},
}

@article{Gross:2017aos,
    author = "Gross, David J. and Rosenhaus, Vladimir",
    title = "{All point correlation functions in SYK}",
    eprint = "1710.08113",
    archivePrefix = "arXiv",
    primaryClass = "hep-th",
    doi = "10.1007/JHEP12(2017)148",
    journal = "JHEP",
    volume = "12",
    pages = "148",
    year = "2017"
}

@article{Bergshoeff:1989ns,
    author = "Bergshoeff, E. and Blencowe, M. P. and Stelle, K. S.",
    title = "{Area Preserving Diffeomorphisms and Higher Spin Algebra}",
    reportNumber = "IMPERIAL/TH/88-89/9",
    doi = "10.1007/BF02108779",
    journal = "Commun. Math. Phys.",
    volume = "128",
    pages = "213",
    year = "1990"
}

@article{Alkalaev:2022szj,
    author = "Alkalaev, Konstantin and Joung, Euihun and Yoon, Junggi",
    title = "{Color decorations of Jackiw-Teitelboim gravity}",
    eprint = "2204.10214",
    archivePrefix = "arXiv",
    primaryClass = "hep-th",
    doi = "10.1007/JHEP08(2022)286",
    journal = "JHEP",
    volume = "08",
    pages = "286",
    year = "2022"
}

@article{Gonzalez:2018enk,
    author = "Gonz{\'a}lez, Hern{\'a}n A. and Grumiller, Daniel and Salzer, Jakob",
    title = "{Towards a bulk description of higher spin SYK}",
    eprint = "1802.01562",
    archivePrefix = "arXiv",
    primaryClass = "hep-th",
    reportNumber = "TUW--18--02",
    doi = "10.1007/JHEP05(2018)083",
    journal = "JHEP",
    volume = "05",
    pages = "083",
    year = "2018"
}

@article{Vasiliev:1995sv,
    author = "Vasiliev, Mikhail A.",
    title = "{Higher spin gauge interactions for matter fields in two-dimensions}",
    eprint = "hep-th/9511063",
    archivePrefix = "arXiv",
    reportNumber = "FIAN-TD-15-95",
    doi = "10.1016/0370-2693(95)01122-7",
    journal = "Phys. Lett. B",
    volume = "363",
    pages = "51--57",
    year = "1995"
}

@article{Gross:2017vhb,
    author = "Gross, David J. and Rosenhaus, Vladimir",
    title = "{A line of CFTs: from generalized free fields to SYK}",
    eprint = "1706.07015",
    archivePrefix = "arXiv",
    primaryClass = "hep-th",
    doi = "10.1007/JHEP07(2017)086",
    journal = "JHEP",
    volume = "07",
    pages = "086",
    year = "2017"
}

@article{Anninos:2023lin,
    author = {Anninos, Dionysios and Anous, Tarek and Pethybridge, Ben and {\c{S}}eng{\"o}r, Gizem},
    title = "{The discreet charm of the discrete series in dS$_{2}$}",
    eprint = "2307.15832",
    archivePrefix = "arXiv",
    primaryClass = "hep-th",
    doi = "10.1088/1751-8121/ad14ad",
    journal = "J. Phys. A",
    volume = "57",
    number = "2",
    pages = "025401",
    year = "2024"
}

@article{Giombi:2011kc,
    author = "Giombi, Simone and Minwalla, Shiraz and Prakash, Shiroman and Trivedi, Sandip P. and Wadia, Spenta R. and Yin, Xi",
    title = "{Chern-Simons Theory with Vector Fermion Matter}",
    eprint = "1110.4386",
    archivePrefix = "arXiv",
    primaryClass = "hep-th",
    doi = "10.1140/epjc/s10052-012-2112-0",
    journal = "Eur. Phys. J. C",
    volume = "72",
    pages = "2112",
    year = "2012"
}

@article{Leigh:2003gk,
    author = "Leigh, Robert G. and Petkou, Anastasios C.",
    title = "{Holography of the N=1 higher spin theory on AdS(4)}",
    eprint = "hep-th/0304217",
    archivePrefix = "arXiv",
    reportNumber = "CERN-TH-2003-095",
    doi = "10.1088/1126-6708/2003/06/011",
    journal = "JHEP",
    volume = "06",
    pages = "011",
    year = "2003"
}

@article{Sezgin:2003pt,
    author = "Sezgin, E. and Sundell, P.",
    title = "{Holography in 4D (super) higher spin theories and a test via cubic scalar couplings}",
    eprint = "hep-th/0305040",
    archivePrefix = "arXiv",
    reportNumber = "MIFP-03-09, UU-07-03",
    doi = "10.1088/1126-6708/2005/07/044",
    journal = "JHEP",
    volume = "07",
    pages = "044",
    year = "2005"
}

@article{Blencowe:1988gj,
    author = "Blencowe, M. P.",
    title = "{A Consistent Interacting Massless Higher Spin Field Theory in $D$ = (2+1)}",
    reportNumber = "IMPERIAL/TP-87-88/30",
    doi = "10.1088/0264-9381/6/4/005",
    journal = "Class. Quant. Grav.",
    volume = "6",
    pages = "443",
    year = "1989"
}

@article{Gaberdiel:2012uj,
    author = "Gaberdiel, Matthias R. and Gopakumar, Rajesh",
    title = "{Minimal Model Holography}",
    eprint = "1207.6697",
    archivePrefix = "arXiv",
    primaryClass = "hep-th",
    doi = "10.1088/1751-8113/46/21/214002",
    journal = "J. Phys. A",
    volume = "46",
    pages = "214002",
    year = "2013"
}

@article{Callan:1992rs,
    author = "Callan, Jr., Curtis G. and Giddings, Steven B. and Harvey, Jeffrey A. and Strominger, Andrew",
    title = "{Evanescent black holes}",
    eprint = "hep-th/9111056",
    archivePrefix = "arXiv",
    reportNumber = "UCSB-TH-91-54, EFI-91-67, PUPT-1294",
    doi = "10.1103/PhysRevD.45.R1005",
    journal = "Phys. Rev. D",
    volume = "45",
    number = "4",
    pages = "R1005",
    year = "1992"
}

@article{Bekaert:2025azj,
    author = "Bekaert, Xavier and Sharapov, Alexey and Skvortsov, Evgeny",
    title = "{Higher-Spin Poisson Sigma Models and Holographic Duality for SYK Models}",
    eprint = "2509.19964",
    archivePrefix = "arXiv",
    primaryClass = "hep-th",
    month = "9",
    year = "2025"
}

@article{Isler:1989hq,
    author = "Isler, K. and Trugenberger, C. A.",
    title = "{A Gauge Theory of Two-dimensional Quantum Gravity}",
    reportNumber = "MIT-CTP-1739",
    doi = "10.1103/PhysRevLett.63.834",
    journal = "Phys. Rev. Lett.",
    volume = "63",
    pages = "834",
    year = "1989"
}

@article{Jackiw:1992bw,
    author = "Jackiw, R.",
    title = "{Gauge theories for gravity on a line}",
    eprint = "hep-th/9206093",
    archivePrefix = "arXiv",
    reportNumber = "MIT-CTP-2105",
    doi = "10.1007/BF01017075",
    journal = "Theor. Math. Phys.",
    volume = "92",
    pages = "979--987",
    year = "1992"
}

@article{Gonzalez:2013oaa,
    author = "Gonzalez, Hernan A. and Matulich, Javier and Pino, Miguel and Troncoso, Ricardo",
    title = "{Asymptotically flat spacetimes in three-dimensional higher spin gravity}",
    eprint = "1307.5651",
    archivePrefix = "arXiv",
    primaryClass = "hep-th",
    reportNumber = "CECS-PHY-13-06",
    doi = "10.1007/JHEP09(2013)016",
    journal = "JHEP",
    volume = "09",
    pages = "016",
    year = "2013"
}

@article{Afshar:2013vka,
    author = "Afshar, Hamid and Bagchi, Arjun and Fareghbal, Reza and Grumiller, Daniel and Rosseel, Jan",
    title = "{Spin-3 Gravity in Three-Dimensional Flat Space}",
    eprint = "1307.4768",
    archivePrefix = "arXiv",
    primaryClass = "hep-th",
    reportNumber = "TUW-13-09",
    doi = "10.1103/PhysRevLett.111.121603",
    journal = "Phys. Rev. Lett.",
    volume = "111",
    number = "12",
    pages = "121603",
    year = "2013"
}

@article{Grumiller:2020elf,
    author = "Grumiller, Daniel and Hartong, Jelle and Prohazka, Stefan and Salzer, Jakob",
    title = "{Limits of JT gravity}",
    eprint = "2011.13870",
    archivePrefix = "arXiv",
    primaryClass = "hep-th",
    reportNumber = "TUW--20--05",
    doi = "10.1007/JHEP02(2021)134",
    journal = "JHEP",
    volume = "02",
    pages = "134",
    year = "2021"
}

@article{Schrader:1972zd,
    author = "Schrader, R.",
    title = "{The maxwell group and the quantum theory of particles in classical homogeneous electromagnetic fields}",
    doi = "10.1002/prop.19720201202",
    journal = "Fortsch. Phys.",
    volume = "20",
    pages = "701--734",
    year = "1972"
}

@article{Bonanos:2008ez,
    author = "Bonanos, Sotirios and Gomis, Joaquim",
    title = "{Infinite Sequence of Poincare Group Extensions: Structure and Dynamics}",
    eprint = "0812.4140",
    archivePrefix = "arXiv",
    primaryClass = "hep-th",
    reportNumber = "ICCUB-08-147, KEK-1292, UB-ECM-PF-08-23",
    doi = "10.1088/1751-8113/43/1/015201",
    journal = "J. Phys. A",
    volume = "43",
    pages = "015201",
    year = "2010"
}

@article{Gomis:2017cmt,
    author = "Gomis, Joaquim and Kleinschmidt, Axel",
    title = "{On free Lie algebras and particles in electro-magnetic fields}",
    eprint = "1705.05854",
    archivePrefix = "arXiv",
    primaryClass = "hep-th",
    reportNumber = "ICCUB-17--002",
    doi = "10.1007/JHEP07(2017)085",
    journal = "JHEP",
    volume = "07",
    pages = "085",
    year = "2017"
}

@article{Cangemi:1992bj,
    author = "Cangemi, D. and Jackiw, R.",
    title = "{Gauge invariant formulations of lineal gravity}",
    eprint = "hep-th/9203056",
    archivePrefix = "arXiv",
    reportNumber = "MIT-CTP-2085",
    doi = "10.1103/PhysRevLett.69.233",
    journal = "Phys. Rev. Lett.",
    volume = "69",
    pages = "233--236",
    year = "1992"
}

@article{Campoleoni:2021blr,
    author = "Campoleoni, Andrea and Pekar, Simon",
    title = "{Carrollian and Galilean conformal higher-spin algebras in any dimensions}",
    eprint = "2110.07794",
    archivePrefix = "arXiv",
    primaryClass = "hep-th",
    doi = "10.1007/JHEP02(2022)150",
    journal = "JHEP",
    volume = "02",
    pages = "150",
    year = "2022"
}

@article{Alkalaev:2020kut,
    author = "Alkalaev, Konstantin and Bekaert, Xavier",
    title = "{On BF-type higher-spin actions in two dimensions}",
    eprint = "2002.02387",
    archivePrefix = "arXiv",
    primaryClass = "hep-th",
    doi = "10.1007/JHEP05(2020)158",
    journal = "JHEP",
    volume = "05",
    pages = "158",
    year = "2020"
}

@article{Alkalaev:2019xuv,
    author = "Alkalaev, Konstantin and Bekaert, Xavier",
    title = "{Towards higher-spin AdS$_2$/CFT$_1$ holography}",
    eprint = "1911.13212",
    archivePrefix = "arXiv",
    primaryClass = "hep-th",
    doi = "10.1007/JHEP04(2020)206",
    journal = "JHEP",
    volume = "04",
    pages = "206",
    year = "2020"
}

@article{Prokushkin:1998bq,
      author         = "Prokushkin, S. F. and Vasiliev, Mikhail A.",
      title          = "{Higher spin gauge interactions for massive matter fields
                        in 3-D AdS space-time}",
      journal        = "Nucl. Phys.",
      volume         = "B545",
      year           = "1999",
      pages          = "385",
      doi            = "10.1016/S0550-3213(98)00839-6",
      eprint         = "hep-th/9806236",
      archivePrefix  = "arXiv",
      primaryClass   = "hep-th",
      reportNumber   = "FIAN-TD-16-98",
      SLACcitation   = "%%CITATION = HEP-TH/9806236;%%"
}

@article{Chamseddine:1989wn,
    author = "Chamseddine, Ali H. and Wyler, D.",
    title = "{Topological Gravity in (1+1)-dimensions}",
    reportNumber = "PRINT-89-0605 (ZURICH)",
    doi = "10.1016/0550-3213(90)90460-U",
    journal = "Nucl. Phys. B",
    volume = "340",
    pages = "595--616",
    year = "1990"
}

@book{Arnold89,
    author="V.I. Arnol'd",
    title="Mathematical Methods of Classical Mechanics",
    publisher="Springer-Verlag New York",
    year="1989",
    doi="10.1007/978-1-4757-2063-1"
}

@article{Sharapov:2019vyd,
      author         = "Sharapov, Alexey and Skvortsov, Evgeny",
      title          = "{Formal Higher Spin Gravities}",
      journal        = "Nucl. Phys.",
      volume         = "B941",
      year           = "2019",
      pages          = "838-860",
      doi            = "10.1016/j.nuclphysb.2019.02.011",
      eprint         = "1901.01426",
      archivePrefix  = "arXiv",
      primaryClass   = "hep-th",
      SLACcitation   = "%%CITATION = ARXIV:1901.01426;%%"
}

@article{Jackiw:1984je,
      author         = "Jackiw, R.",
      title          = "{Lower Dimensional Gravity}",
      booktitle      = "{1984 Meeting of the Division of Particles and Fields of
                        the APS Santa Fe, New Mexico, October 31-November 3,
                        1984}",
      journal        = "Nucl. Phys.",
      volume         = "B252",
      year           = "1985",
      pages          = "343-356",
      doi            = "10.1016/0550-3213(85)90448-1",
      reportNumber   = "MIT-CTP-1203",
      SLACcitation   = "%%CITATION = NUPHA,B252,343;%%"
}

@article{Teitelboim:1983ux,
      author         = "Teitelboim, C.",
      title          = "{Gravitation and Hamiltonian Structure in Two Space-Time
                        Dimensions}",
      journal        = "Phys. Lett.",
      volume         = "126B",
      year           = "1983",
      pages          = "41-45",
      doi            = "10.1016/0370-2693(83)90012-6",
      SLACcitation   = "%%CITATION = PHLTA,126B,41;%%"
}

@article{Fukuyama:1985gg,
      author         = "Fukuyama, Takeshi and Kamimura, Kiyoshi",
      title          = "{Gauge Theory of Two-dimensional Gravity}",
      journal        = "Phys. Lett.",
      volume         = "160B",
      year           = "1985",
      pages          = "259-262",
      doi            = "10.1016/0370-2693(85)91322-X",
      reportNumber   = "Print-85-0318 (TOHO)",
      SLACcitation   = "%%CITATION = PHLTA,160B,259;%%"
}

@article{Feigin,
      author         = "Feigin, Boris L.",
      title          = "{Lie algebras $gl(\lambda)$ and cohomologies of Lie algebras of differential operators}",
      journal        = "Russian Math. Surv.",
      volume         = "43",
      pages          = "169-170",
      doi            = "",
      year           = "1988",
      reportNumber   = "",
      SLACcitation   = "",
}

@article{Antunes:2025iaw,
    author = "Antunes, Ant{\'o}nio and Levine, Nat and Meineri, Marco",
    title = "{Demystifying integrable QFTs in AdS: No-go theorems for higher-spin charges}",
    eprint = "2502.06937",
    archivePrefix = "arXiv",
    primaryClass = "hep-th",
    doi = "10.21468/SciPostPhys.20.3.088",
    journal = "SciPost Phys.",
    volume = "20",
    pages = "088",
    year = "2026"
}

@article{Bordemann:1989zi,
    author = "Bordemann, M. and Hoppe, J. and Schaller, P.",
    title = "{Infinite-dimensional matrix algebras}",
    reportNumber = "KA-THEP-10-1989",
    doi = "10.1016/0370-2693(89)91687-0",
    journal = "Phys. Lett. B",
    volume = "232",
    pages = "199--203",
    year = "1989"
}

@article{Fradkin:1990ir,
      author         = "Fradkin, E. S. and Linetsky, V. {\relax Ya}.",
      title          = "{Infinite dimensional generalizations of simple Lie
                        algebras}",
      journal        = "Mod. Phys. Lett.",
      volume         = "A5",
      year           = "1990",
      pages          = "1967-1977",
      doi            = "10.1142/S0217732390002249",
      SLACcitation   = "%%CITATION = MPLAE,A5,1967;%%"
}

@article{Klebanov:2002ja,
      author         = "Klebanov, I. R. and Polyakov, A. M.",
      title          = "{AdS dual of the critical O(N) vector model}",
      journal        = "Phys. Lett.",
      volume         = "B550",
      year           = "2002",
      pages          = "213-219",
      doi            = "10.1016/S0370-2693(02)02980-5",
      eprint         = "hep-th/0210114",
      archivePrefix  = "arXiv",
      primaryClass   = "hep-th",
      reportNumber   = "PUPT-2053",
      SLACcitation   = "%%CITATION = HEP-TH/0210114;%%"
}

@inproceedings{Prokushkin:1998vn,
    author = "Prokushkin, Sergey and Vasiliev, Mikhail A.",
    title = "{3-d higher spin gauge theories with matter}",
    booktitle = "{2nd International Seminar on Supersymmetries and Quantum Symmetries}: {Dedicated to the Memory of Victor I. Ogievetsky}",
    eprint = "hep-th/9812242",
    archivePrefix = "arXiv",
    reportNumber = "FIAN-TD-27-98",
    month = "12",
    year = "1998"
}

@article{Ammon:2017vwt,
    author = "Ammon, Martin and Grumiller, Daniel and Prohazka, Stefan and Riegler, Max and Wutte, Raphaela",
    title = "{Higher-Spin Flat Space Cosmologies with Soft Hair}",
    eprint = "1703.02594",
    archivePrefix = "arXiv",
    primaryClass = "hep-th",
    reportNumber = "TUW--17--01",
    doi = "10.1007/JHEP05(2017)031",
    journal = "JHEP",
    volume = "05",
    pages = "031",
    year = "2017"
}

@article{Alkalaev:2014qpa,
      author         = "Alkalaev, K. B.",
      title          = "{Global and local properties of AdS$_{2}$ higher spin
                        gravity}",
      journal        = "JHEP",
      volume         = "10",
      year           = "2014",
      pages          = "122",
      doi            = "10.1007/JHEP10(2014)122",
      eprint         = "1404.5330",
      archivePrefix  = "arXiv",
      primaryClass   = "hep-th",
      SLACcitation   = "%%CITATION = ARXIV:1404.5330;%%"
}

@article{Alkalaev:2013fsa,
      author         = "Alkalaev, K. B.",
      title          = "{On higher spin extension of the Jackiw-Teitelboim
                        gravity model}",
      journal        = "J. Phys.",
      volume         = "A47",
      year           = "2014",
      pages          = "365401",
      doi            = "10.1088/1751-8113/47/36/365401",
      eprint         = "1311.5119",
      archivePrefix  = "arXiv",
      primaryClass   = "hep-th",
      reportNumber   = "FIAN-TD-2013-10",
      SLACcitation   = "%%CITATION = ARXIV:1311.5119;%%"
}

@ARTICLE{Campoleoni:2010zq,
  author = {Campoleoni, Andrea and Fredenhagen, Stefan and Pfenninger, Stefan
    and Theisen, Stefan},
  title = {{Asymptotic symmetries of three-dimensional gravity coupled to higher-spin
    fields}},
  journal = {JHEP},
  year = {2010},
  volume = {11},
  pages = {007},
  archiveprefix = {arXiv},
  doi = {10.1007/JHEP11(2010)007},
  eprint = {1008.4744},
  primaryclass = {hep-th},
  slaccitation = {%%CITATION = 1008.4744;%%}
}

@article{Gaberdiel:2010pz,
      author         = "Gaberdiel, Matthias R. and Gopakumar, Rajesh",
      title          = "{An AdS Dual for Minimal Model CFTs}",
      journal        = "Phys. Rev.",
      volume         = "D83",
      year           = "2011",
      pages          = "066007",
      doi            = "10.1103/PhysRevD.83.066007",
      eprint         = "1011.2986",
      archivePrefix  = "arXiv",
      primaryClass   = "hep-th",
      SLACcitation   = "%%CITATION = ARXIV:1011.2986;%%"
}

@ARTICLE{Henneaux:2010xg,
  author = {Henneaux, Marc and Rey, Soo-Jong},
  title = {{Nonlinear W(infinity) Algebra as Asymptotic Symmetry of Three-Dimensional
    Higher Spin Anti-de Sitter Gravity}},
  year = {2010},
  archiveprefix = {arXiv},
  eprint = {1008.4579},
  primaryclass = {hep-th},
  slaccitation = {%%CITATION = 1008.4579;%%}
}

@article{Afshar:2019axx,
    author = "Afshar, Hamid and Gonz\'alez, Hern\'an A. and Grumiller, Daniel and Vassilevich, Dmitri",
    title = "{Flat space holography and the complex Sachdev-Ye-Kitaev model}",
    eprint = "1911.05739",
    archivePrefix = "arXiv",
    primaryClass = "hep-th",
    reportNumber = "TUW-19-04",
    doi = "10.1103/PhysRevD.101.086024",
    journal = "Phys. Rev. D",
    volume = "101",
    number = "8",
    pages = "086024",
    year = "2020"
}

@article{Afshar:2021qvi,
    author = "Afshar, Hamid and Oblak, Blagoje",
    title = "{Flat JT gravity and the BMS-Schwarzian}",
    eprint = "2112.14609",
    archivePrefix = "arXiv",
    primaryClass = "hep-th",
    doi = "10.1007/JHEP11(2022)172",
    journal = "JHEP",
    volume = "11",
    pages = "172",
    year = "2022"
}

@article{Ammon:2020fxs,
    author = "Ammon, Martin and Pannier, Michel and Riegler, Max",
    title = "{Scalar Fields in 3D Asymptotically Flat Higher-Spin Gravity}",
    eprint = "2009.14210",
    archivePrefix = "arXiv",
    primaryClass = "hep-th",
    doi = "10.1088/1751-8121/abdbc6",
    journal = "J. Phys. A",
    volume = "54",
    number = "10",
    pages = "105401",
    year = "2021"
}

@article{Ammon:2022vjr,
    author = "Ammon, Martin and Pannier, Michel",
    title = "{Unfolded Fierz-Pauli equations in three-dimensional asymptotically flat spacetimes}",
    eprint = "2211.12530",
    archivePrefix = "arXiv",
    primaryClass = "hep-th",
    doi = "10.1007/JHEP02(2023)161",
    journal = "JHEP",
    volume = "02",
    pages = "161",
    year = "2023"
}

@article{Cardenas:2018krd,
    author = "C{\'a}rdenas, Marcela and Fuentealba, Oscar and Gonz{\'a}lez, Hern{\'a}n A. and Grumiller, Daniel and Valc{\'a}rcel, Carlos and Vassilevich, Dmitri",
    title = "{Boundary theories for dilaton supergravity in 2D}",
    eprint = "1809.07208",
    archivePrefix = "arXiv",
    primaryClass = "hep-th",
    reportNumber = "TUW-18-05",
    doi = "10.1007/JHEP11(2018)077",
    journal = "JHEP",
    volume = "11",
    pages = "077",
    year = "2018"
}

@article{Montano:1990ru,
    author = "Montano, David and Aoki, Kenichiro and Sonnenschein, Jacob",
    title = "{Topological Supergravity in Two-dimensions}",
    reportNumber = "UCLA-90-TEP-22, CALT-68-1637",
    doi = "10.1016/0370-2693(90)91050-L",
    journal = "Phys. Lett. B",
    volume = "247",
    pages = "64--70",
    year = "1990"
}

@article{Griguolo:2023aem,
    author = "Griguolo, Luca and Guerrini, Luigi and Panerai, Rodolfo and Papalini, Jacopo and Seminara, Domenico",
    title = "{Supersymmetric localization of (higher-spin) JT gravity: a bulk perspective}",
    eprint = "2307.01274",
    archivePrefix = "arXiv",
    primaryClass = "hep-th",
    doi = "10.1007/JHEP12(2023)124",
    journal = "JHEP",
    volume = "12",
    pages = "124",
    year = "2023"
}

@article{Kruthoff:2022voq,
    author = "Kruthoff, Jorrit",
    title = "{Higher spin JT gravity and a matrix model dual}",
    eprint = "2204.09685",
    archivePrefix = "arXiv",
    primaryClass = "hep-th",
    doi = "10.1007/JHEP09(2022)017",
    journal = "JHEP",
    volume = "09",
    pages = "017",
    year = "2022"
}

@article{Campoleoni:2016vsh,
    author = "Campoleoni, Andrea and Gonzalez, Hernan A. and Oblak, Blagoje and Riegler, Max",
    editor = "Brink, Lars and Henneaux, Marc and Vasiliev, Mikhail A.",
    title = "{BMS Modules in Three Dimensions}",
    eprint = "1603.03812",
    archivePrefix = "arXiv",
    primaryClass = "hep-th",
    doi = "10.1142/S0217751X16500688",
    journal = "Int. J. Mod. Phys. A",
    volume = "31",
    number = "12",
    pages = "1650068",
    year = "2016"
}

@article{Campoleoni:2015qrh,
    author = "Campoleoni, Andrea and Gonzalez, Hernan A. and Oblak, Blagoje and Riegler, Max",
    title = "{Rotating Higher Spin Partition Functions and Extended BMS Symmetries}",
    eprint = "1512.03353",
    archivePrefix = "arXiv",
    primaryClass = "hep-th",
    doi = "10.1007/JHEP04(2016)034",
    journal = "JHEP",
    volume = "04",
    pages = "034",
    year = "2016"
}

@article{Grumiller:2014lna,
    author = "Grumiller, D. and Riegler, M. and Rosseel, J.",
    title = "{Unitarity in three-dimensional flat space higher spin theories}",
    eprint = "1403.5297",
    archivePrefix = "arXiv",
    primaryClass = "hep-th",
    reportNumber = "TUW-14-05",
    doi = "10.1007/JHEP07(2014)015",
    journal = "JHEP",
    volume = "07",
    pages = "015",
    year = "2014"
}

@article{Grumiller:2013swa,
    author = "Grumiller, Daniel and Leston, Mauricio and Vassilevich, Dmitri",
    title = "{Anti-de Sitter holography for gravity and higher spin theories in two dimensions}",
    eprint = "1311.7413",
    archivePrefix = "arXiv",
    primaryClass = "hep-th",
    reportNumber = "TUW-13-18",
    doi = "10.1103/PhysRevD.89.044001",
    journal = "Phys. Rev. D",
    volume = "89",
    number = "4",
    pages = "044001",
    year = "2014"
}

@article{Teitelboim:1983uy,
    author = "Teitelboim, C.",
    title = "{Supergravity and Hamiltonian structure in two space-time dimensions}",
    doi = "10.1016/0370-2693(83)90013-8",
    journal = "Phys. Lett. B",
    volume = "126",
    pages = "46--48",
    year = "1983"
}

@article{Pope:1989sr,
    author = "Pope, C. N. and Romans, L. J. and Shen, X.",
    title = "{$W$(infinity) and the Racah-wigner Algebra}",
    reportNumber = "USC-89-HEP040, CTP-TAMU-72-89",
    doi = "10.1016/0550-3213(90)90539-P",
    journal = "Nucl. Phys. B",
    volume = "339",
    pages = "191--221",
    year = "1990"
}
\end{document}